\documentclass[review]{elsarticle}

\biboptions{sort&compress}

\usepackage[utf8x]{inputenc}
\usepackage[T1]{fontenc}
\usepackage[polish,english]{babel}

\usepackage{amsmath,amsfonts} 
\usepackage{hyperref}
\usepackage[x11names]{xcolor}
\usepackage{xspace}

\newcommand{\TRP}{\textrm{TRP}}
\newcommand{\dd}{\mathrm{d}}
\newcommand{\e}{\mathrm{e}}

\newcommand{\etal}{\textit{et~al.}\xspace}
\newcommand{\eg}{\textit{e.g.}}

\begin{document}

\begin{frontmatter}

\title{A~mathematical model of low grade gliomas treated with temozolomide and its therapeutical implications}

 \author[uw,uclm]{M.~U.~Bogda\'nska\corref{cor1}}
    \ead{m.bogdanska@mimuw.edu.pl}
 
\author[uw]{M.~Bodnar}
    \ead{m.bodnar@mimuw.edu.pl}

\author[uclm]{J.~Belmonte-Beitia}
    \ead{juan.belmonte@uclm.es}

\author[bern]{M.~Murek}
    \ead{michael.murek@insel.ch}
    
\author[bern]{P.~Schucht}
    \ead{philippe.schucht@insel.ch}

\author[bern]{J.~Beck}
    \ead{juergen.beck@insel.ch}
                 
\author[uclm]{V.~M.~P\'erez-Garc\'ia}
    \ead{victor.perezgarcia@uclm.es}
 
 \cortext[cor1]{Corresponding author} 
 
\address[uw]{Faculty of Mathematics, Informatics and Mechanics, University of Warsaw, \\
ul. Banacha 2, 02-097 Warsaw, Poland}
\address[uclm]{Departamento de Matem\'aticas, Universidad de Castilla-La Mancha, \\ 
ETSI~Industriales, Avda. Camilo Jos\'e Cela 3, 13071 Ciudad Real, Spain} 
\address[bern]{Universit\"atsklinik f\"ur Neurochirurgie, Bern University Hospital, \\
CH-3010 Bern, Switzerland}

\begin{abstract}
Low grade gliomas (LGGs) are infiltrative and incurable primary brain tumours with typically slow evolution. These tumours usually occur in young and otherwise healthy patients, bringing controversies in treatment planning since aggressive treatment may lead to undesirable side effects. Thus, for management decisions it would be valuable to obtain early estimates of LGG growth potential.
 
Here we propose a~simple mathematical model of LGG growth and its response to chemotherapy which allows the growth of LGGs to be described in real patients. The model predicts, and our clinical data confirms, that the speed of response to chemotherapy is related to tumour aggressiveness. Moreover, we provide a~formula for the time to radiological progression, which can be possibly used as a~measure of tumour aggressiveness. 
 
Finally, we suggest that the response to a~few chemotherapy cycles upon diagnosis might be used to predict tumour growth and to guide therapeutical actions on the basis of the findings.
\end{abstract}

\end{frontmatter}

\section{Introduction}
Low grade gliomas (LGGs) are slowly growing tumours which arise from supporting glial cells in the brain. Although they proliferate slowly, the ultimate behaviour of these tumours is not benign. They are almost invariably incurable due to their diffusive infiltrative nature and often lead to patient death due to malignant transformation, \emph{i.e.} transformation of the tumour into more aggressive anaplastic form.

Treatment of~LGGs is controversial among clinicians as LGGs usually occur in young and otherwise healthy patients. Besides seizures, patients with LGGs appear neurologically asymptomatic. Thus, life-prolonging treatment should not come at the cost of compromising the quality of life of these otherwise healthy and mostly young patients. Management decisions, whether a~patient with LGG should receive resection, radiation therapy or chemotherapy, are not fully standardized.
Clinicians usually base their decisions about extent of surgical resection, timing of radiotherapy and long-term benefits and risks of chemotherapy on a large number of factors including age, performance status, location of tumour and patient preference. 

Current standard of care is first to perform maximal resection with minimum impact on the brain's functional areas. Support for such an approach comes from studies which demonstrate that the extent of resection is a~prognostic factor for LGG patients \cite{Keles,Smith}. Unfortunately, because of the LGGs infiltrative characteristics, surgery alone fails to cure these tumours in most cases as only the tumour bulk can be resected. There is a~trend to use more active treatments and various alternative approaches have been considered. Post-operative radiotherapy could be a~therapeutic option for low-grade gliomas, but it causes long-term neurocognitive toxicity and fails to show a significant improvement in patient survival~\cite{Karim,Karim1,Shaw}. Therefore radiotherapy is usually deferred and performed routinely only in patients with tumours facing a~high risk of malignant transformation \cite{Pouratian}. 

In this context, there is an increasing interest in the use of chemotherapeutic agents which could influence tumour evolution and at the same time allow the delay of more aggressive treatments. Currently there are  two chemotherapeutic drugs effective for the treatment of glioma patients: temozolomide (TMZ), an oral alkylating agent and procarbazine, lomustine and vincristine (PCV), a~combination of alkylating agents and cell-cycle specific microtubule inhibitor. About 25-50\% of LGGs show chemotherapeutic responses to treatment with either TMZ~or PCV. PCV has been used from the 1970's and has been suggested as a~clinically relevant option especially for some subtypes of anaplastic gliomas with the 1p/19q codeletion \cite{Cairncross, Bent2}. Unfortunately it causes significant myelosuppression, nausea and peripheral neuropathy \cite{Mason,Buckner}. 

TMZ~is a~cell-cycle non-specific prodrug, absorbed with almost 100\% bioavailability \cite{Newlands}.~This chemotherapeutic agent can cross the blood-brain barrier and is spontaneously converted in tumour cells to its active metabolite \cite{Marchesi}. The main cytotoxic action of TMZ occurs at the O$^6-$ position of guanine. During subsequent cycles of replication, the futile mismatch repair system is initiated and O$^6$MeG is incorrectly paired with thymine instead of cytosine, eventually inducing cell death, see \emph{e.g.} \cite{Jiricny,Barciszewska}. A~phase III~trial showed the efficacy of TMZ for high-grade gliomas~\cite{Stupp} and since then it has been used routinely for patients with newly diagnosed glioblastoma (the most malignant type of glioma)~\cite{Bent3}. Phase II~trials have demonstrated its effectivity against both previously irradiated and unirradiated LGGs \cite{Kesari,Kaloshi,Pouratian1}. In addition, there are reports of cases where neoadjuvant chemotherapy given to surgically unresectable tumours has allowed subsequent gross total resection \cite{Blonski,Jo}, which is of great importance especially when the tumour is highly infiltrative or located in~eloquent areas. Large prospective studies with long-term follow-up are under development to verify statistically significant improvement in resections following TMZ~treatment. It has also been reported that TMZ~treatment may lead to an important reduction in seizure frequency in LGG patients  \cite{Koekkoek}. Thus prolonged TMZ~treatment until evidence of resistance is a~clinically interesting option for selected patients as an up-front or adjuvant treatment. Clinical trials are on the way to study the effect of this treatment on survival. 

Temozolomide has a~better toxicity profile than PCV, being well tolerated by patients \cite{Liu,Pace,Pouratian1}. This is due to the fact that, in contrast to procarbazine and lomustine, TMZ does not cause a chemical cross-linking of the DNA strands and it is less toxic to the haematopoietic progenitor cells in the bone marrow \cite{Agarwala}. Patients treated with TMZ are also seeing a reduced risk of cumulative haematologic toxicity thanks to the rapid elimination of this drug \cite{Agarwala,Portnow,Newlands}. 

The response of glioma cells to chemotherapy is a~subject of many clinical and biological studies. Recently, it has been shown that in some LGG patients a~metabolic response to TMZ~therapy is observed even 3 months after the end of~the treatment \cite{Wyss,Guillevin}. Moreover, dynamic volumetric studies have shown that a~treatment-related volume decrease can be observed for many months after the chemotherapy is discontinued \cite{Peyre,Ricard}. The time to maximum tumour response was reported to be in some cases larger than 2 years \cite{Chamberlain}.~Other researchers investigated relations between some molecular characteristics of LGGs and response to TMZ~\cite{Hoang,Kaloshi,Ricard, Ohba}. However, the question of the correct timing of chemotherapy remains unanswered, namely whether it should be given first or when progression has been observed. Another issue to be addressed is~the optimal fractioning of TMZ.

The typical plan of TMZ~treatment is to give doses of $150$--$200$~mg per m$^2$~of patient body surface once per day for 5 days every 28 consecutive days. The number of such cycles in~clinical practise is usually between 12 and 30 \cite{Ricard, Ribba,Chamberlain} and it depends on patient-related characteristics and the haematological toxicity observed. There have been many clinical studies on alternative treatment regimens for gliomas. Among others in \cite{Kesari} patients were treated in cycles with doses of 75 mg/m$^2$ given daily for 7~weeks followed by four-week breaks. Some trials on dose escalation  \cite{Wick,Taal,Viaccoz} intended to overcome the DNA-repair activity of the enzyme MGMT (O$^6$MeG methyltransferase), which reverses alkylation at the O$^6$ position of guanine \cite{Everhard,Hegi}. 
However, these TMZ~regimes were either not effective or had to be rejected because of a~high toxicity \cite{Hammond}. Recent studies show that O$^6$MeG is able to trigger not only apoptosis (programmed cell death), but also autophagy (mechanism allowing the degradation and recycling of cellular components) and senescence (state of permanent cell-cycle arrest in~which cells are resistant to apoptotic death) \cite{Knizhnik}, which may provide an explanation why previous studies failed to improve treatment efficacy. Many clinicians conclude that the chemotherapy fractionation scheme providing the best tumour response and acceptable haematologic toxicity are still to be determined.
 
Mathematical modelling has great potential to help in finding appropriate therapeutic timings and/or fractionations and in individual patient treatment decision. Even models simple from mathematical point of view can be useful for clinical practice. 

In this paper we describe a~simple mathematical model for LGG growth and response to chemotherapy that fits very well with longitudinal volumetric data of patients diagnosed with LGGs. Interestingly, the model suggests that the response of the tumour to chemotherapy may be related to~its aggressiveness. We also provide an approximate explicit formula for the time of tumour response to chemotherapy. This equation may be helpful to clinicians in selecting patients who will benefit most from early treatment and finding the best personalised therapy.

Our plan in this paper is as follows: in Section 2 we develop the model and describe its parameters. In Section 3 we present the results of model fitting to patient data and suggest the relation between time of response to chemotherapy and patient prognosis. The analytical estimates are shown in Section 4. We discuss the results of our model and therapeutic implications in Section 5.

\section{Mathematical model} \label{sec:model}
\subsection{The dynamics of tumour cells}
Computational modelling of glioma growth started 20 years ago \cite{Murray} and has received strong attention in the last few years (see \emph{e.g.}  \cite{Unkelbach,Konukoglu,Barazzoul, Bastogne,Kirkby}). 
For solid tumours mathematical models have considered different chemothe\-rapy-related factors such as drug diffusion, uptake/binding, clearance and their effect on cell cycle progression (see \emph{e.g.} \cite{Au,Jackson,Tzafriri,Swierniak}). Many mechanistic mathematical models models have been developed to improve the design of chemotherapy regimes (see \emph{e.g.} \cite{Gardner} for a~summary). However, very few models have considered chemotherapy of LGGs \cite{Ribba}.

In order to keep our description as simple as possible we will build a~continuous macroscopic model assuming that the tumour grows due to net cell division. The simplest choice for the proliferative term is to assume that a~tumour cell number $P(t)$ is governed by a~logistic growth with coefficient $\rho,$~its inverse giving an estimate of the typical cell doubling times. 
 
TMZ is a small molecule and is easily absorbed in the digestive tract with 100\% of bioavailability and a half-life of absorption of 7 minutes \cite{Ostermann}. It should be noted that the intact TMZ molecule easily crosses the blood brain barrier due to its lipophilicity, and is then activated in the brain compartment. It hydrolyses to MTIC, which subsequently undergoes spontaneous breakdown to an inactive metabolite AIC and an active methyldiazonium cation. It is the methyldiazonium ion that causes the damage in patients' DNA, namely it transfers methyl groups to DNA. The most common sites of methylation are the N$^7$ position of guanine (N$^7$-MeG; 60-80\%) followed by the N$^3$ position of adenine (N$^3$-MeA; 10-20\%) and the O$^6$ position of guanine (O6-MeG; 5-10\%)  \cite{Fu}, the last one being responsible for major cytotoxicity.

As to the TMZ effect on tumour cells, we choose here to formulate a macroscopic model. It is known that TMZ-induced damage leads to cell death long after the end of therapy ~\cite{Peyre,Bent1,Ricard,Chamberlain}. It has been verified \emph{in vitro} that the glioma cells death after administration of TMZ is induced most typically in one of the post-treatment cell cycles \cite{Roos} due to futile mismatch repair cycles (see \emph{e.g.} \cite{Marchesi} for a detailed description of this mechanism). This delayed cell death is a key feature of glioma response to chemotherapy.  Thus, in line with this biological evidence we will assume that tumour cells treated with chemotherapy die after a~time $k/\rho$ of the order of mean $k$~effective tumour doubling times. This type of model has been used successfully to describe the effect of radiotherapy on LGGs \cite{Victor,Victor2}.

Then, we will complement the equation for the number of functionally alive tumour cells $P(t)$ with an equation for the evolution of the number of cells irreversibly damaged by chemotherapy $D(t)$. The number of cells damaged by the~drug in a~time unit is considered to be proportional to the concentration of the drug $C(t)$ multiplied by the number of proliferating tumour cells with the rate~$\alpha,$~measuring the influence of TMZ~on cells.
We assume that irreversibly damaged tumour cells try to enter mitosis with the same probability as those active, but die after a mean value of $k$ such attempts, which results in the growth rate $\rho\left(1-\frac{P+D}{K}\right)$ for proliferative cells and the death rate $-\frac{\rho}{k}\left(1-\frac{P+D}{K}\right)$ for damaged cells, $K$~being the carrying capacity for both populations $P+D$, which leads to the following set of equations
\begin{subequations}
 \label{ode}
	\begin{eqnarray} 
		\frac{\dd P}{\dd t}  & = & \rho P\left(1-\frac{P+D}{K}\right)  - \alpha PC, \label{ode1} \\ 
		\frac{\dd D}{\dd t}  & = & -\frac{\rho}{k}D\left(1-\frac{P+D}{K}\right)  + \alpha PC, \label{ode2} 
	\end{eqnarray}
where $C(t)$ accounts for the brain chemotherapy concentration to be described in more details later. Taking an average cell volume, we can easily treat the tumour mass $P+D$ as the total tumour volume, which is easier to compare with results obtained from magnetic resonance imaging (MRI), usually used in brain tumour diagnosis and follow-up observation.

\subsection{Kinetics of chemotherapy drug} \label{subs:CT}
The systemic pharmacokinetic and pharmacodynamic properties of TMZ has been studied in detail in several studies \cite{Baker,Ostermann}. Baker \etal \cite{Baker} described the concentrations of TMZ,  MTIC and AIC in plasma in detail. Ostermann \etal~\cite{Ostermann} collected data on TMZ concentration from blood and cerebrospinal fluid (CSF) obtained via lumbar puncture in patients with malignant gliomas. 

However due to the physiological separation of brain and tumour from both blood and CSF (through blood-brain, blood-tumour, blood-CSF and CSF-tumour barriers) the amount of drug reaching the tumour differs from the amount of drug circulating in blood and CSF \cite{Shannon,Blakeley}. 
Therefore instead of describing a~complicated mechanism with many unknown parameters based on data collected from blood or CSF, we chose a simpler dynamics based directly on the brain tissue data. Thus we will base our model on data from the study by Portnow \etal~\cite{Portnow} who examined TMZ concentration in intracerebral microdialysis samples from peritumoural brain interstitium obtained from patients with central nervous system tumours.

With regard to chemotherapy pharmacokinetics, we will assume, as usual \cite{Jacqmin,Ribba}, that the concentration of TMZ, $C(t)$, measured in units of days decays exponentially due to the drug clearance with a~constant rate $\lambda$. It is consistent with the fact that TMZ has linear pharmacokinetics \cite{Newlands}. Thus, we have to complement Eqs.~\eqref{ode} with 
\begin{equation} \label{ode3}
		\frac{\dd C}{\dd t}  =  -\lambda C. 
	\end{equation}
\end{subequations}
TMZ~reaches a~maximal drug concentration in the brain about two hours after administration~\cite{Hammond,Portnow,Baker}, what is very short in comparison to the time scale of tumour evolution in the model (of the order of years). Thus, we may treat the whole time of oral drug administration, absorption and transport to the brain as a~discontinuous  change in the drug concentration that occurs at given administration times. Such a formulation of the problem enables also the following mathematical analysis and estimations.

Chemotherapy will consist of a~sequence of doses $d_1,d_2,\ldots,d_n$ given at times 
$t_1 < t_2 <\ldots < t_n$. 
We assume that initially (at time $t=t_0$, taken to be the start of the tumour observation) the tumour has a~certain mass $P_0=P(t_0)$ and there are neither damaged cells, nor chemotherapy drugs within the tumour, thus \mbox{$D(t_0) = 0, \ C(t_0) = 0.$}
Thus for $t\in (0, t_1)$ solving Eqs.~\eqref{ode} with initial conditions given above leads to 
\begin{equation} \label{solP}
P(t) = \frac{ KP_0 \e^{\rho t}}{K - P_0 \left(1-\e^{\rho t} \right)}, \quad D(t) = 0, \quad C(t)=0.
\end{equation}
For $t\in (t_j, t_{j+1})$, $j=1,2,3,\ldots,n$ with $n$ being the total number of doses, the values of $P$, $D$ and $C$ change according to  Eqs.~\eqref{ode}, while at $t=t_j$ subsequent doses of the drug are administrated which we model as impulses, so that  
\begin{equation} \label{impulses}
 P(t_j)=P(t_j^-), \quad 
	 D(t_j)=D(t_j^-), \quad 
	 C(t_j)=C(t_j^-)+C_j, 
\end{equation}
where $f(t_j^-) = \lim_{t\to t_j^-} f(t)$ and $C_{j}$ is the fraction of the dose $d_{j}$ which reaches the tumour tissue, accounting for drug loss during transport to the brain.

The interval between doses (typically 1 day) will be chosen to be larger than the time of whole dose elimination (reported to be around 7h \cite{Agarwala}) and the typical damage repair times (of the order of a~few hours \cite{VanderKogel}), so that one dose will not alter the effect of the next one. 

The asymptotic behaviour of model~\eqref{ode} under the effect of a~finite number of doses is easy to obtain. When no drug is given for time $t \ge t_n$ (with $n$ being the index of the last dosis), then \mbox{$C(t)=C(t_n) \exp(-\lambda (t-t_n))\to 0$} as $t\to+\infty$. As $C \to \infty$ and $P$ is bounded, then $\alpha P C \to 0$ for $t \to +\infty.$ As a consequence $D'(t)<0$ for sufficiently large time. 
Then it is easily seen that $D(t)\to 0$ and $P(t)\to K$ as $t\to+\infty$. This behaviour is not surprising and can be interpreted as patient death due to drug clearance and tumour regrowth. Patient death usually occurs when the tumour reaches a~critical size called the fatal tumour burden considered to be in high-grade glioma models to be around 6 cm in diameter \cite{Swanson2008,Woodward}.

It is important to emphasise that model \eqref{ode} intends to describe the effect of first-line chemotherapy, since after the treatment resistant phenotypes arise leading to the acquisition of drug resistance. Thus a~detailed analysis of second-line chemotherapy would require the introduction of more phenotypes in the model and is beyond the scope of this research.

\subsection{Parameter estimation} \label{sec:parameters}
To work with Eqs. \eqref{ode} we need to provide realistic values for the model parameters. 

The saturation coefficient $K$ for LGG growth will be set to the volume of a~sphere of diameter $10$~cm reported to be the maximal mean tumour diameter observed in LGG patients \cite{Ricard}. In fact, most patients die when the tumour diameter is about 6 cm in size as discussed before \cite{Swanson2008,Woodward}.

We can estimate the rate of drug decay $\lambda$ using values of TMZ~half-life clearance time $t_{1/2}$. From the definition of $t_{1/2}$ and assuming exponential decay as in Eqs.\eqref{ode} we have
$$\frac{1}{2}=\e^{\textstyle -\lambda t_{1/2}}.$$

To account also for the drug loss during transport to the brain we calculate value $C_{j}$ of the maximal dose $d_{j}$ reaching the tumour as 
\begin{equation} \label{dose}
C_j = \beta \cdot d_{j} \cdot b,
\end{equation}
where  $\beta$ is the fraction of TMZ~getting to 1ml of brain interstitial fluid (from a~unit dose) and $b$~is a~surface of a~patient body with $j \in \{1,\ldots, n\}$ and $n$~being the total number of doses $d$ administered. Then $C_{j}$ can be interpreted as an effective dose per fraction. 

The most typical chemotherapy schedule consists of cycles of 28 days with five TMZ~oral doses on days 1 to 5 followed by a~break of 23~days. Standard dose per day is 150 mg per m$^2$ of patient body surface, which is usually around 1.6~m$^2$ for women and 1.9 m$^2$ for men \cite{Mosteller} with an average of 1.7~m$^2$~\cite{Sparreboom}. 
Then in the case of  the standard chemotherapy scheme we will fix $d_{j}=d=150$mg/m$^2$ and $C_{j}=C_0=\beta \cdot d \cdot b.$

The parameter $\beta$ can be calculated using the value of maximal TMZ~concentration $C_{max}$ for a~dose of 150 mg/m$^2$ taken from the literature \cite{Hammond,Portnow}. Assuming that time to reach peak drug concentration in the brain is negligible (equals $0.85-2$h) in comparison to the time scale of the model, we set the initial drug concentration $C_{0}$ in the moment of its administration to the value $C_{max}.$ 

A~summary of the biological parameter values is presented in Table~\ref{table-parameters}. 
\begin{table}[ht] \small 
	\begin{center}
		\caption{Biological parameters describing TMZ concentration in brain.}
		\begin{tabular}{l l l c}
			\hline
			Parameter & Description & Value, references & \\ 
			\hline
			$t_{1/2}$ & TMZ~half-life clearance time & $\simeq 2$h & \cite{Hammond} \\
			$C_{max}$ & mean peak TMZ~concentration & $0.6 \mu$g/ml  & \cite{Portnow} \\
			& in brain interstitium & &\\
			$\lambda$ &  rate of decay of TMZ~& 0.3466/h & Calculated \\
			& & & from \cite{Hammond,Portnow} \\
			$\beta$ & fraction of TMZ~getting & $2.1\cdot 10^{-6}$/ml (m)& Estimated\\ 
			& to brain interstitium &  $2.5 \cdot 10^{-6}$/ml (w)& from \cite{Hammond,Portnow}  \\ 
			\hline
		\end{tabular}
		\label{table-parameters}
	\end{center}
\end{table} 

\section{Numerical results}
\subsection{Model fitting to patient data} \label{sec:model fitting}
To test if our simple model given by Eqs.~\eqref{ode} is able to reflect the dynamics of LGG response to chemotherapy, we have used the model to describe volumetric longitudinal data of patients followed at the Bern University Hospital between 1990 and 2013. In this study we selected data on 18 patients who had been treated with TMZ~out of a total number of 82 LGGs patients, see Table~\ref{tab:patients data}. Radiological glioma growth was quantified by manual measurements of tumour diameters on successive MRIs. The three largest tumour diameters ($D_1,\ D_2, \ D_3$) according to three reference orthogonal planes (axial, coronal and sagittal planes) have been measured and tumour volumes have been estimated using the ellipsoidal approximation: $V=(D_1 \cdot D_2 \cdot D_3)/2$, following the standard clinical practice  \cite{Pallud,Mandonnet}. 

The inclusion criteria for patients for the purpose of model fitting in this study included: (i) biopsy/surgery confirmed LGG (astrocytoma, oligoastrocytoma or oligodendroglioma), (ii) availability of at least 2 MRIs before the onset of TMZ~treatment, (iii) no other treatment given in the period of study and (iv) availability of at least 4 MRIs after TMZ~onset with at least one after the end of the chemotherapy. 7 patients satisfied these criteria. All patients in this group received more than 4~TMZ~cycles and the mean duration of TMZ~treatment was 6.26 months. 

\begin{table}[ht] \small 
\begin{center}
\caption{Characteristics of patients treated with TMZ}
\begin{tabular}{l c c}
\hline
Age at diagnosis, mean (st.~deviation), yr & 47.19  (7.54)  \\
Sex, M/F & 14/4  \\
\emph{Histology at diagnosis} &  \\
\hspace{0.5cm} Oligodendroglioma & 7  \\
\hspace{0.5cm} Oligoastrocytoma & 9  \\
\hspace{0.5cm} Astrocytoma & 1  \\
\hspace{0.5cm} Unknown & 1   \\
\emph{Type of surgery} &   \\
\hspace{0.5cm} Biopsy & 10 \\
\hspace{0.5cm} Resection & 9 \\
\emph{Radiotherapy} & 8  \\
\emph{Chemotheraphy (CT)} & 18   \\
\hspace{0.5cm} Age at CT onset, mean (st.~deviation), yr & 51.8 (8.35) \\
\hspace{0.5cm} Time from surgery to CT, mean (st.~deviation), yr & 3.7 (4) \\ 
\hspace{0.5cm} Second-line CT & 8  \\
\hline
\end{tabular}
\label{tab:patients data}
\end{center}
\end{table}

The rate of tumour cell proliferation $\rho,$~the coefficient $k$ describing the delay in damaged cell death and the parameter of TMZ-cell kill strength $\alpha$ were considered to be tumour-specific and fitted for each patient. Thus only three parameters are unknown and the others are taken as in Table \ref{table-parameters}.
 
To estimate the parameter $\rho$ we used patient data before the start of TMZ~administration since it is the only relevant parameter during that time, see Eq.~\eqref{solP}. The initial tumour volume $P_0$in this equation was fixed to be the volume from first MRI~done after surgery. Then, having obtained the value of $\rho$,  MRI~data after the onset of chemotherapy was used to estimate parameters $\alpha$~and $k$ in Eqs.~\eqref{ode}. Model fitting was done using a~weighted least squares method. To simulate Eqs.~\eqref{ode}, we have used the standard Matlab ODE solver based on the Runge-Kutta 4th-order method.

\begin{figure}[h!p]  
\begin{center}
\includegraphics[width=0.525\textwidth]{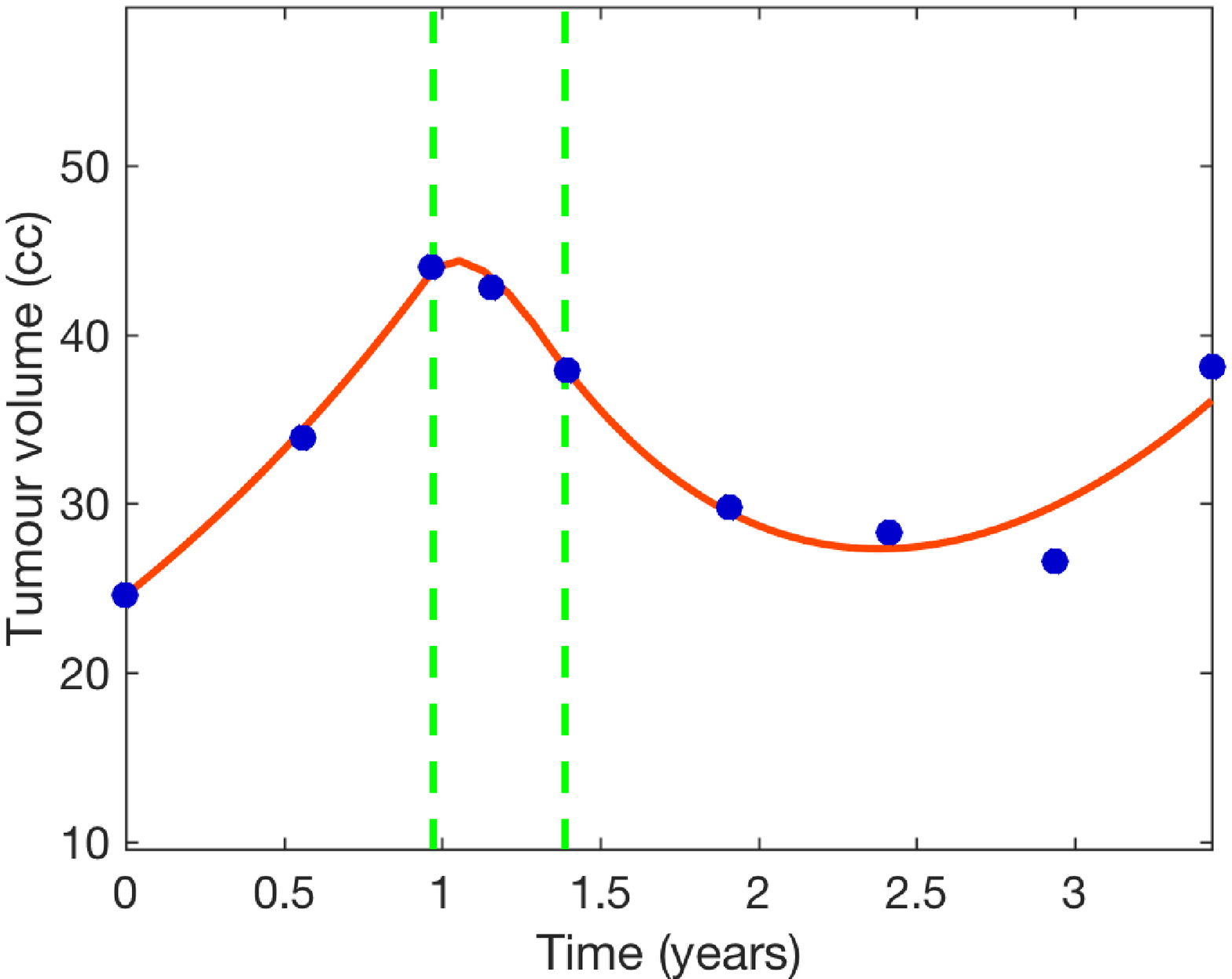} 
\includegraphics[width=0.525\textwidth]{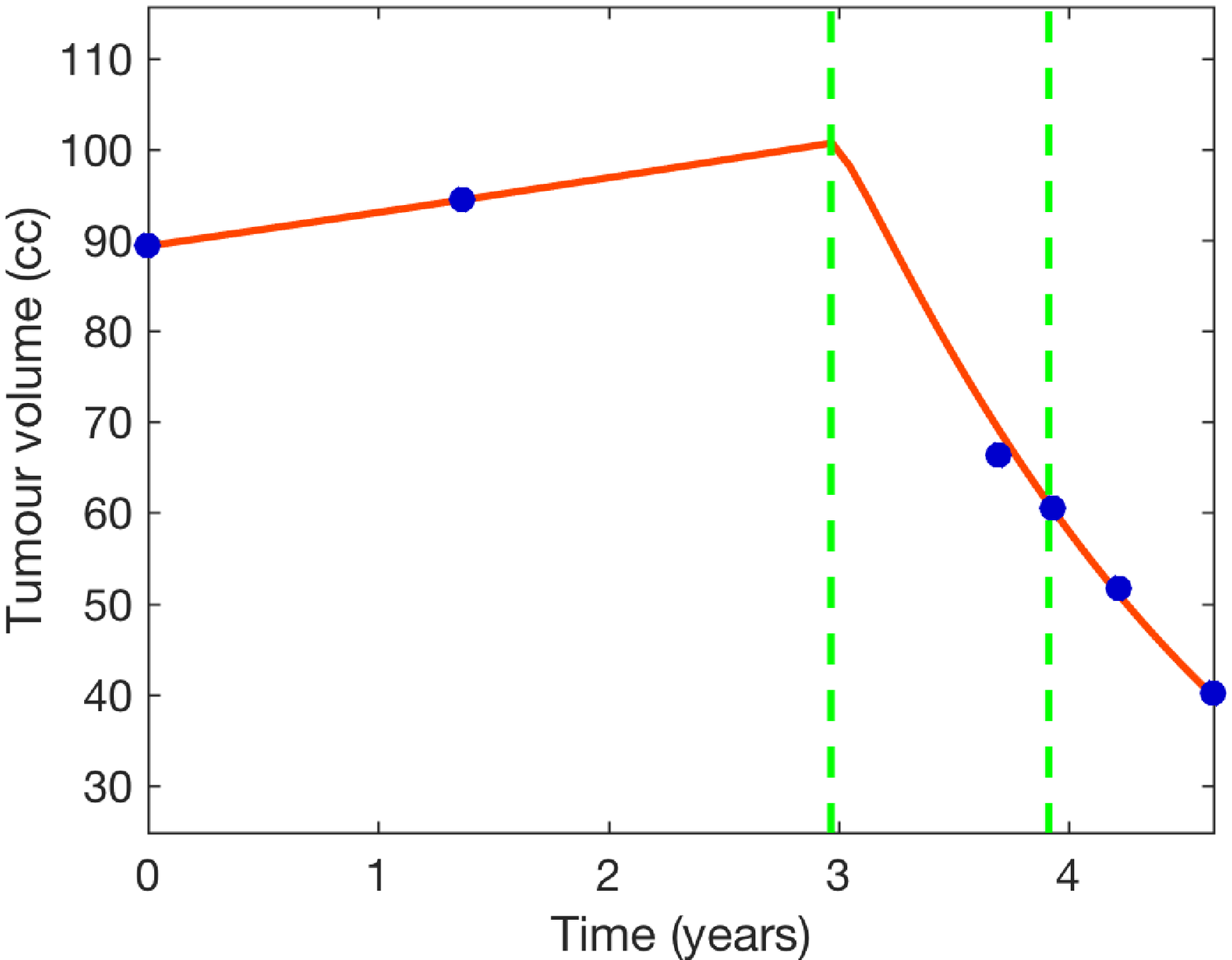} 
\includegraphics[width=0.525\textwidth]{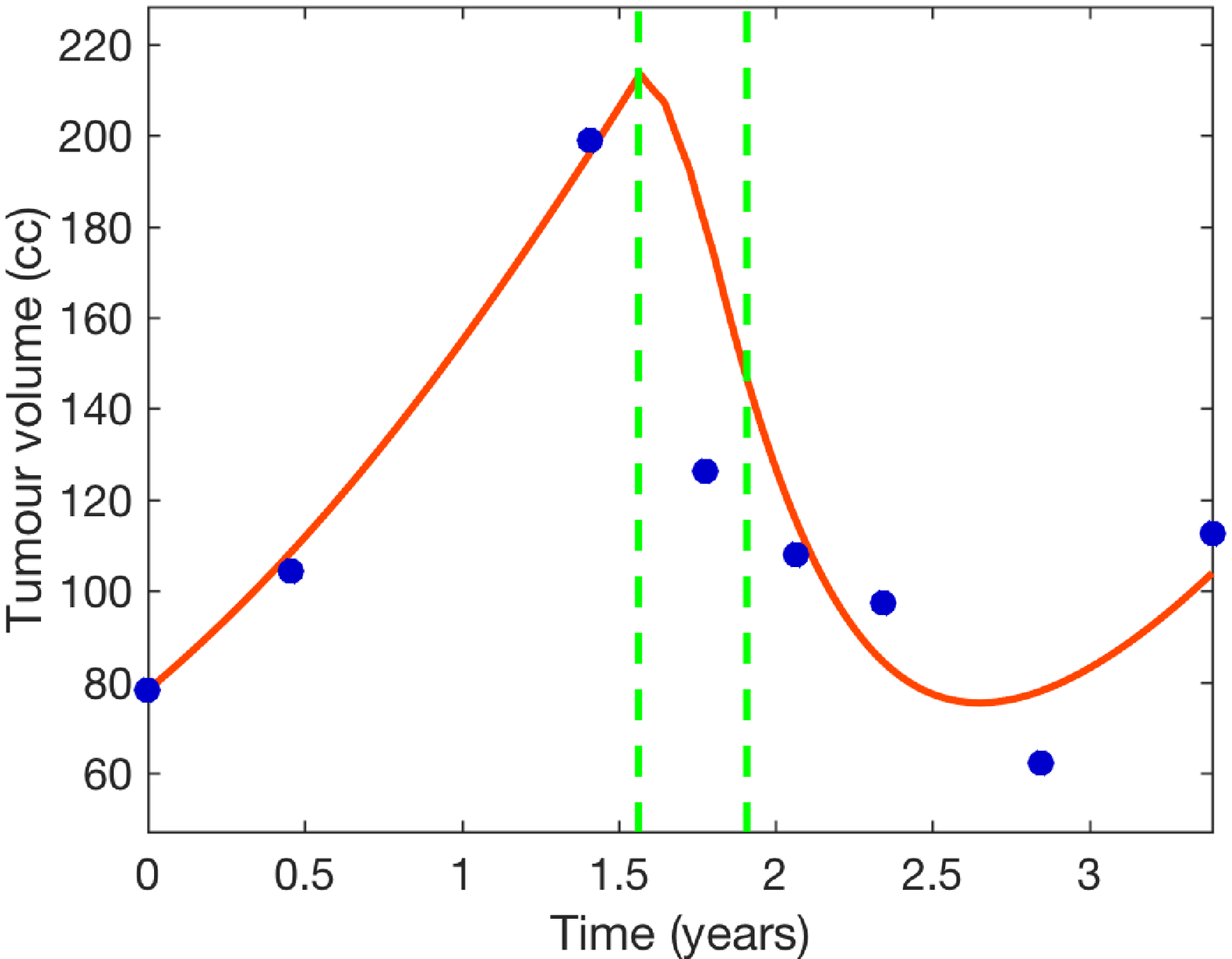} 
\caption{Tumour volume evolution for three selected patients treated with TMZ. The beginning and the end of TMZ~treatment are marked with vertical dashed lines. There are shown the volumes calculated from MRIs (circles) and from the fitted mathematical model (solid lines). The number of TMZ~cycles and the values of parameters were different for each patient.
(top) Patient treated with 5 TMZ~cycles,  $\alpha=0.971918$ml/$\mu$g/day, $\rho = 0.001761$/day, $ \ k = 0.555867.$
(center) Patient treated with 11 TMZ~cycles, \mbox{$\alpha = 0.279911$ml/$\mu$g/day,} $\rho = 0.000136$/day, $ \ k = 0.025617.$ 
(bottom) Patient treated with 4 TMZ~cycles, $\alpha=1.387798$ml/$\mu$g/day, \mbox{$\rho = 0.002416$/day,} $ \ k = 0.272291.$}
\label{fig:proliferacja}
\end{center}
\end{figure}

Figs.~\ref{fig:proliferacja}, \ref{fig:tTTP} show both the real tumour volume data obtained from the MRIs (circles) together with the best fit obtained with Eqs.~\eqref{ode} (solid line). The model dynamics fit the real volumetric tumour evolution well, showing an impressive agreement with a~minimal number of parameters for patients with delayed response to chemotherapy. The minimal value of the fitted proliferation rate for some patients is one order of magnitude smaller than values $(1-5) \cdot 10^{-3}$ day$^{-1}$ observed in other studies  \cite{Gerin,Victor,Victor3} as in these studies the model for tumour growth also considered a diffusive term. Some of the tumours were relatively large, however no formation of neoangiogenesis or necrotic core was observed. See also Supplementary material for all patients data and estimated parameters.

\subsection{Tumours with faster response have worse prognosis} \label{sec:relations}
\begin{figure}[h!t]
\begin{center}
\includegraphics[trim={0.5cm 0cm 1.8cm 0.9cm},clip=true,width=0.45\textwidth]{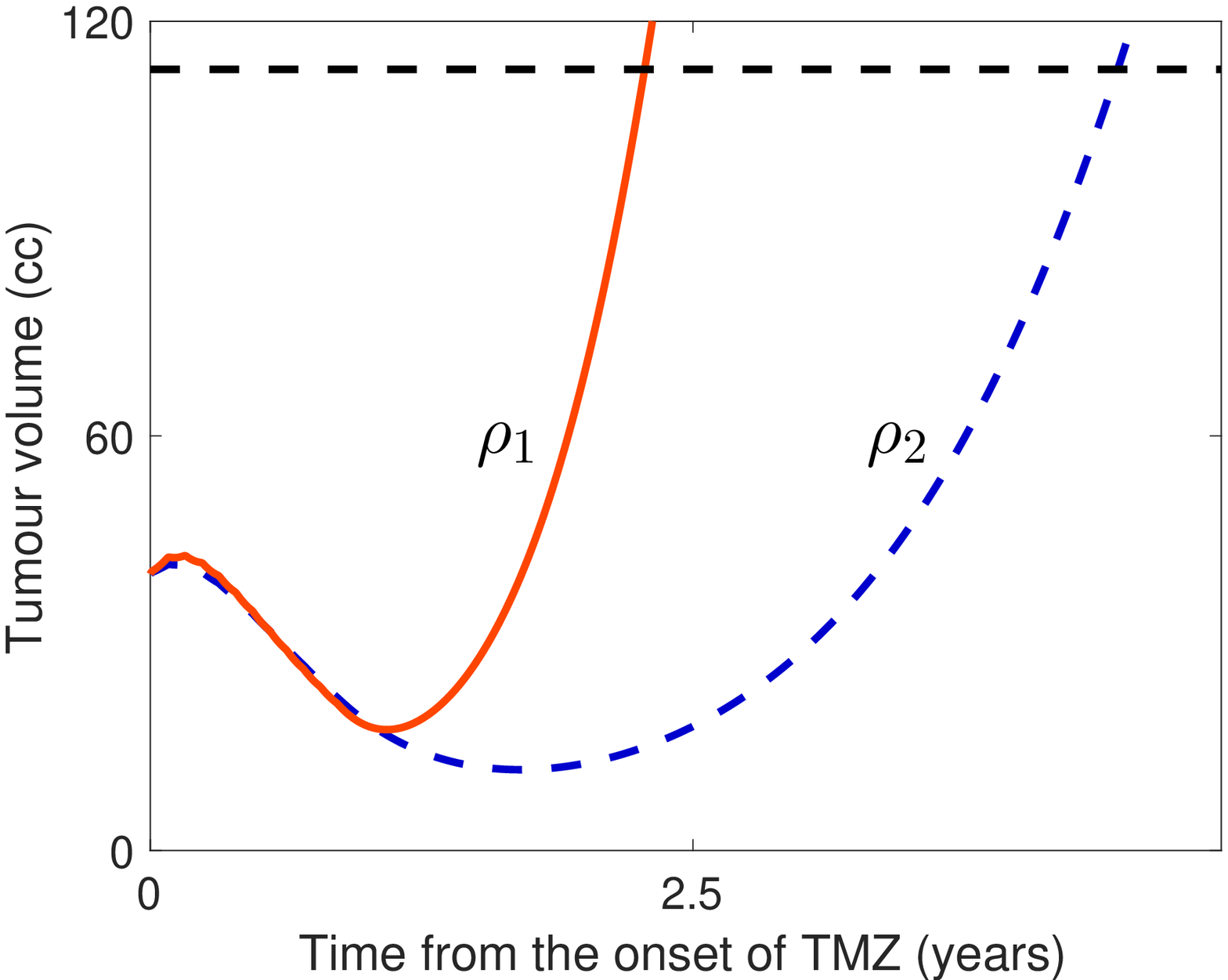}
\includegraphics[trim={0.5cm 0cm 1.8cm 0.9cm},clip=true,width=0.45\textwidth]{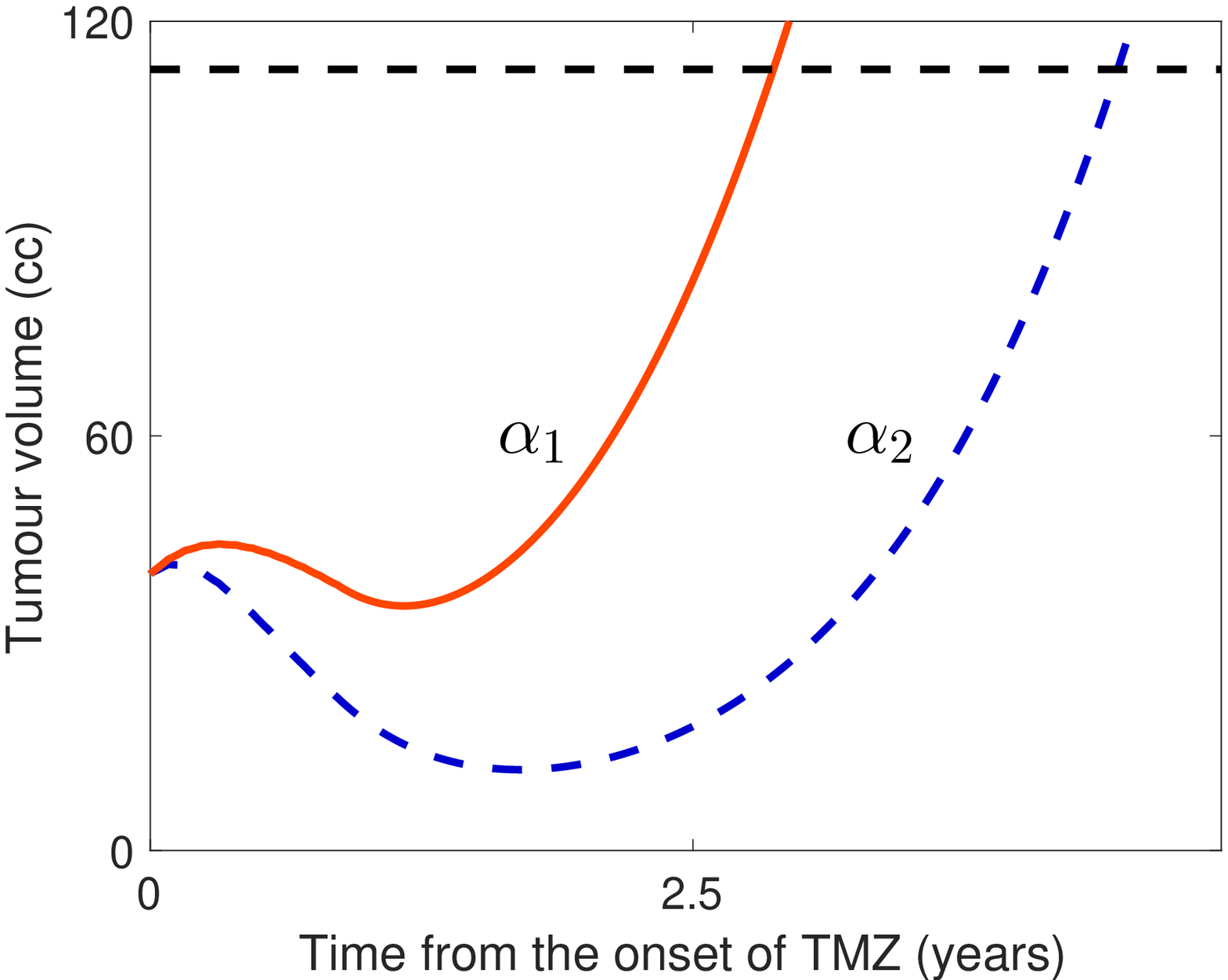}
\caption{Tumour volume evolution for different values of parameters. Virtual patients were treated with 12 cycles of TMZ~as in the  standard fractionation scheme (see Sec.\ref{sec:parameters}) with $k=0.5.$ (left)~Parameter $\alpha$ was fixed to value 0.7ml/$\mu$g/day, $\rho_1=0.004/$day$, \ \rho_2=0.002/$day. (right) Parameter $\rho$ was fixed to value 0.0025/day, 
$\alpha_1=0.4$ml/$\mu$g/day,  $\alpha_2=0.8$ml/$\mu$g/day. The horizontal dotted lines correspond to tumour sizes equal to the fatal tumour burden.}
\label{fig_rho,alpha}
\end{center}
\end{figure}

\begin{figure}[h!t]
\centering
\includegraphics[trim={0.865cm 0cm 1.73cm 0.9cm},clip=true,width=0.49\textwidth]{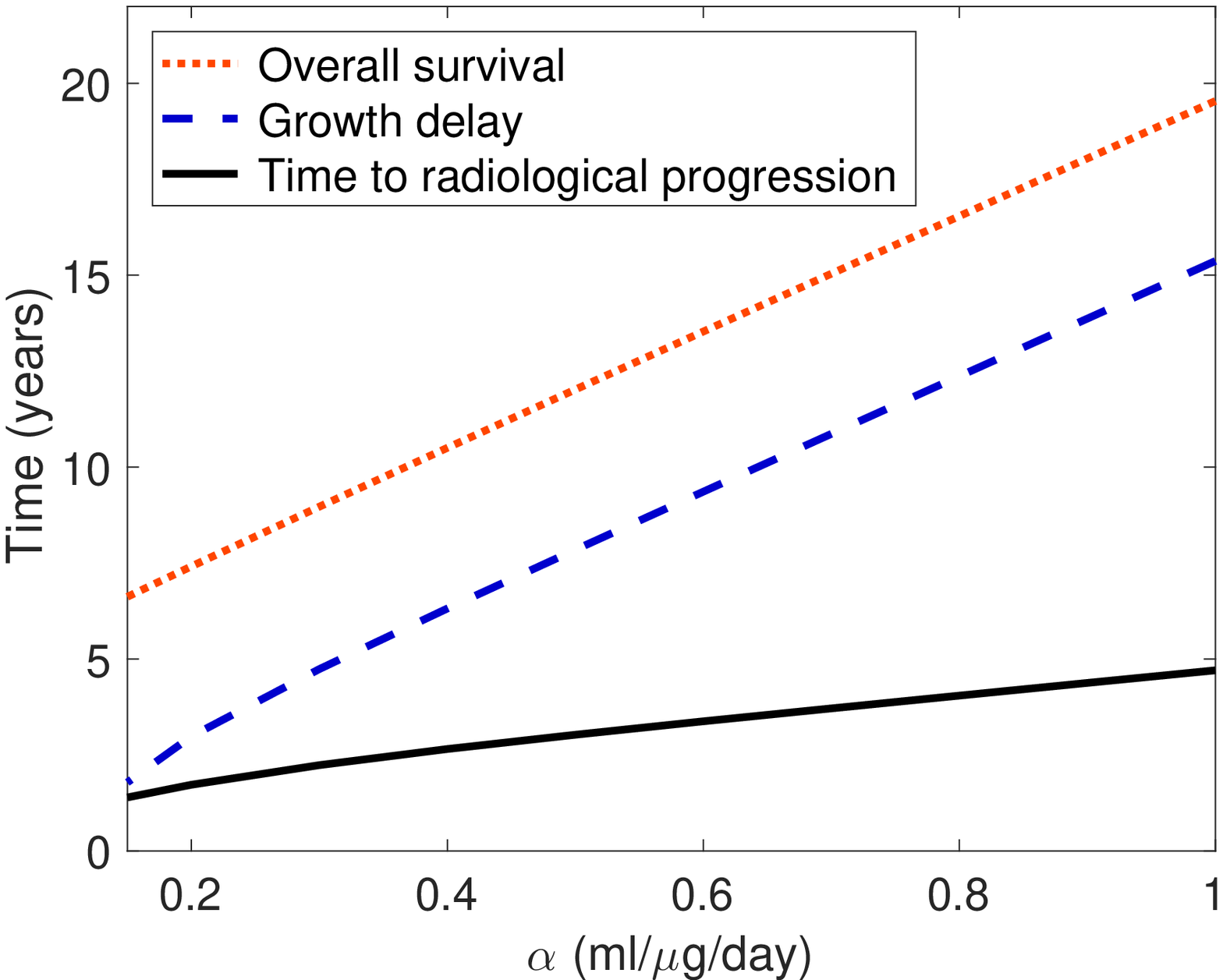} 
\includegraphics[trim={0.865cm 0cm 1.73cm 0.9cm},clip=true,width=0.49\textwidth]{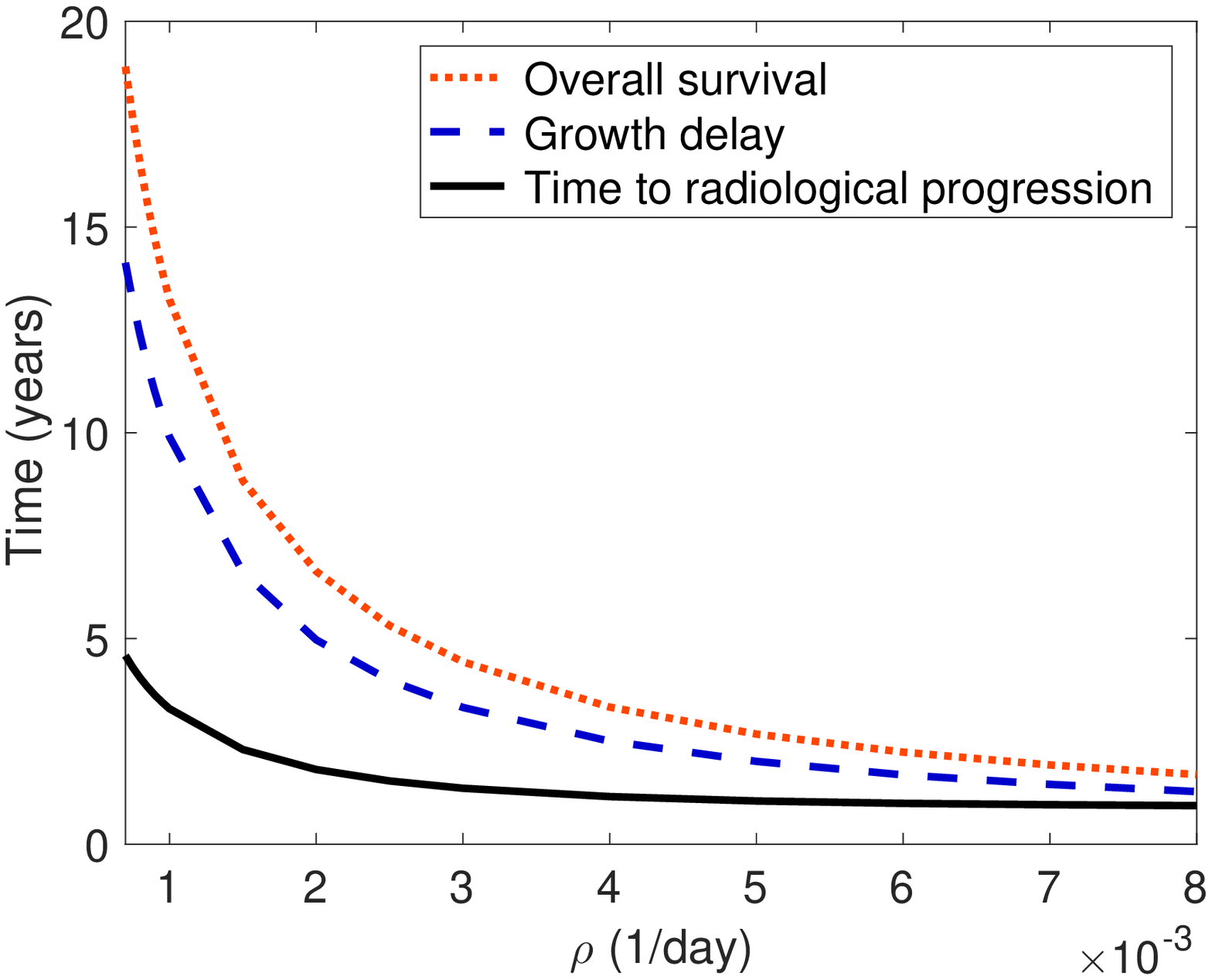}
\caption{Characteristic times of tumour response for different proliferation rates $\rho$ and different levels of TMZ~cell kill strength $\alpha$.~We considered 12 cycles of TMZ~as in the standard fractionation scheme (see Sec.~\ref{sec:parameters}) for virtual patients with LGG of initial volume 40 cm$^3$. (left) Results for $\rho = 0.0008$/day, $k=0.3$ and $\alpha \in [0.1,1]$ml/$\mu$g/day.~(right) Results for $\alpha=0.8$ml/$\mu$g/day, $k=0.3$ and $\rho \in [0.7,8] \times 10^{-3}$/day.}
\label{fig:relacje}
\end{figure}

\begin{figure}[h!t]
\centering
\includegraphics[trim={1cm 0.6cm 1.03cm 1.4cm},clip=true,width=0.47\textwidth]{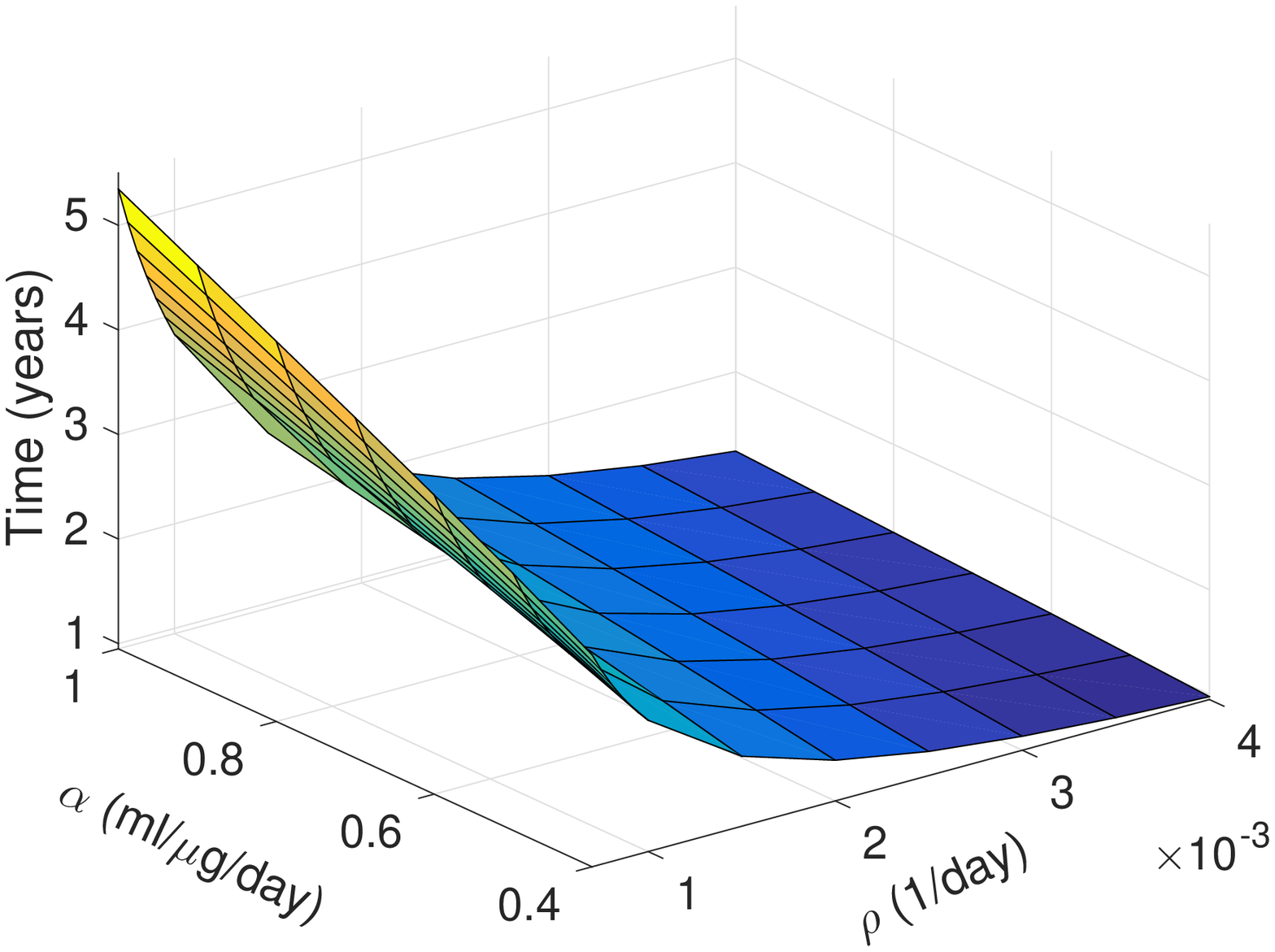} 
\includegraphics[trim={0.63cm 0.6cm 0.95cm 1.4cm},clip=true,width=0.47\textwidth]{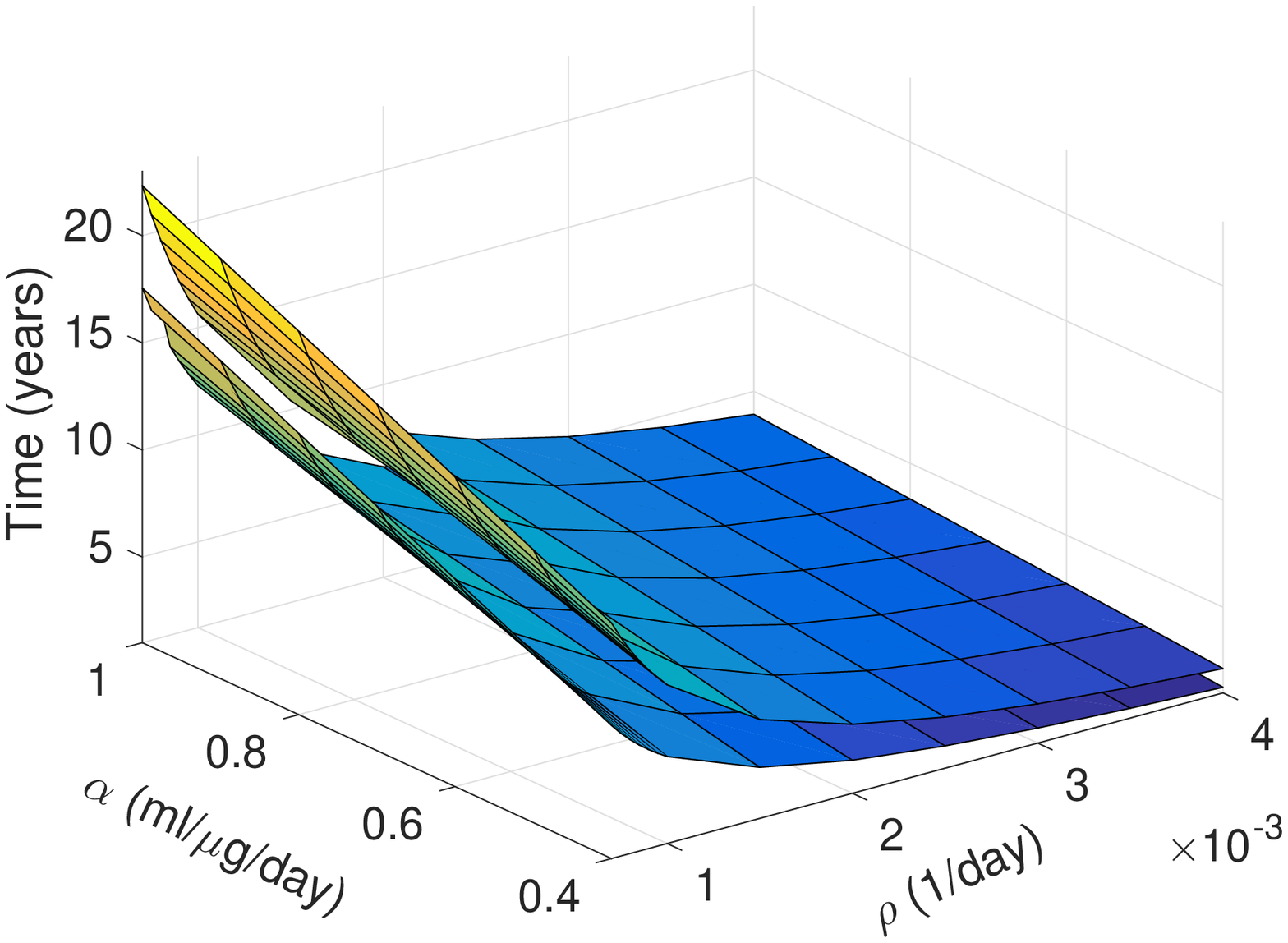}
\caption{Characteristic times of tumour response for different proliferation rates $\rho$ and different levels of TMZ~cell kill strength $\alpha$.~We considered 12 cycles of TMZ~as in the standard fractionation scheme (see Sec.~\ref{sec:parameters}) with $k=0.3.$
Values of time to radiological progression (left), growth delay and overall survival (right) are shown for virtual patients with LGG of an initial volume of 40~cm$^3$.
}
\label{fig:surf}
\end{figure}

\begin{figure}[h!t]
\centering 
\includegraphics[trim={0.83cm 0cm 1.8cm 1.02cm},clip=true,width=0.5\textwidth]{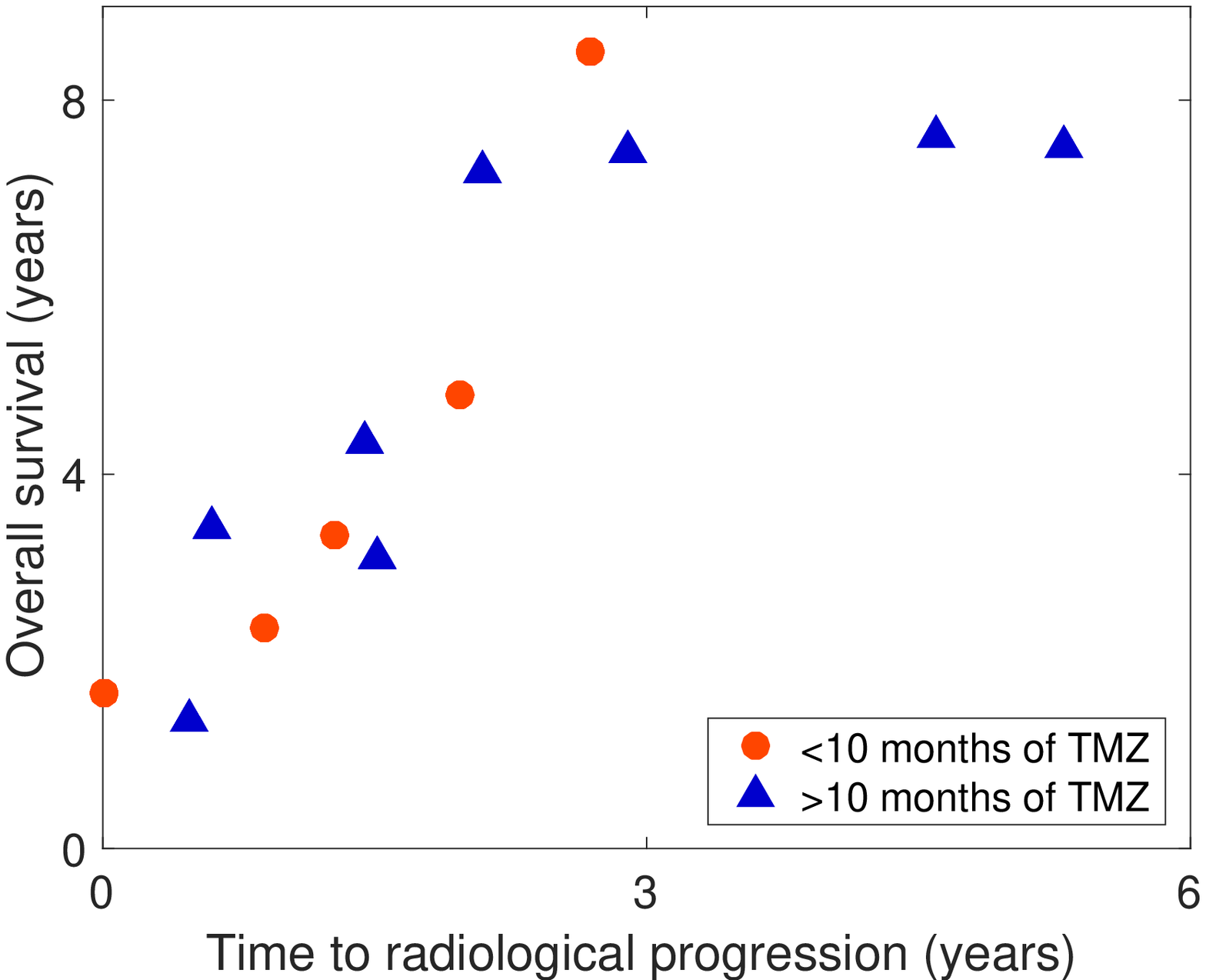}
\caption{Time to radiological response and overall survival of LGGs patients treated with TMZ. Patients were divided into two groups: in the first, denoted by circles, patients were treated with TMZ~for less than 10 months (average: 5.72,  st.~deviation: 3.8), in the second group, denoted by triangles, patients were treated with TMZ~for more than 10 months (average: 13.65,  st.~deviation: 3.14).}
\label{fig:patients-PFS,OS}
\end{figure}
We have also studied how the tumour response depends on parameters fitted, see Figs. \ref{fig:relacje}, \ref{fig:surf}. 
We will denote by ``time to radiological progression'' (TRP) the time when the tumour attains its minimum volume after the chemotherapy onset and starts regrowing. We will refer to ``growth delay'' as the time for which the tumour volume equals the initial one when regrowing after the therapy, see~\cite{Victor}. We will refer to ``early response'' when TRP is attained shortly after the end of chemotherapy and ``no response'' when there was no decrease in tumour volume detectable by radiologist. In case of frequent MRIs these times can be easily obtained from model simulations and compared with the values obtained from patient's MRI~volumetry. It can be estimated with an error being the time between two subsequent MRIs. We have also computed the overall survival (OS) as the time until reaching the fatal tumour burden defined in Sec.\ref{subs:CT}. We have considered only the cases of virtual tumours whose volumes decreased below the volume at TMZ onset. Note that for the purpose of analysis of response to TMZ, OS is computed for virtual tumours responding to TMZ and without any other treatment in the following course of disease, thus in general could be overestimated and thus should not be compared to the values of real-patients overall survival. 

After performing many simulations for different initial values and chemotherapy schemes, we conclude that both larger proliferation rates $\rho$ and smaller TMZ~cell kill strengths $\alpha$, lead to an earlier response to TMZ~treatment. A~more systematic study is shown in Fig.~\ref{fig_rho,alpha} for two specific parameter sets, providing representative examples. Virtual patients who responded earlier to TMZ~(had smaller TRP) had a~faster regrowth and reached the fatal tumour burden earlier. Thus, a shorter TRP is an indicator of worse prognosis. 

Are these model features also present in the patient's data? Fig.~\ref{fig:patients-PFS,OS} shows how the overall survival rate correlates with the time to radiological progression. For the purpose of this analysis we also included patients with only one MRI before treatment with TMZ. We excluded (i) two patients in which only two MRIs were available after the end of chemotherapy showing no tumour regrowth and (ii) one patient treated with TMZ~for only 1.5 month showing no response. For each patient TRP was estimated as the time to the MRI in which tumour volume was the smallest after the onset of treatment. Due to the differing duration of TMZ treatments we divided patients into a group receiving less than 10 TMZ cycles and those receiving 10 or more cycles.

The Spearman rank correlation coefficient between TRP and OS equals 1 for data of patients treated with TMZ~for less than 10 months and 0.9047619 for those treated with TMZ~for longer time. The exact Spearman coefficient test significance levels equal 0.008333 and 0.002282 for right-tailed tests for group of patients treated with less and more than 10 cycles of TMZ, respectively.  This result indicates a~positive correlation between TRP and OS. The significance levels were calculated using R. 
Data on overall survival is right-censored, however the results suggest that the early regrowth of the tumour after chemotherapy is related to its aggressiveness. Despite therapies used after progression, those tumours that responded faster to TMZ~treatment progressed faster, suggesting either larger proliferation potential and/or smaller TMZ~cell kill strength. 

\section{Results (II): Analytical estimates of tumour response} 

\subsection{Survival fraction}
Up to now our analysis has been based on numerical simulations of Eqs. \eqref{ode}. We can calculate the fraction of tumour cells eliminated by a~single dose of chemotherapy, which in the context of radiotherapy is usually referred to as \emph{survival fraction}. To do so, we will assume that the time of drug absorption, distribution and elimination from the human body is much shorter than the doubling time of the tumour cell population, what is true for LGGs. Therefore, focusing on short-term effects of the drug we may neglect the term describing tumour proliferation. Consequently the size of damaged cells population remains zero and we consider instead the~simplified model
\begin{subequations}
\begin{eqnarray}
		\frac{d P}{d t}  & = - \alpha P C, \\ 
		\frac{d C}{d t}  & = -\lambda C.
\end{eqnarray}
	\end{subequations}
For time before the second drug administration $t_1\le t<t_2$ we have
\begin{subequations}
\begin{align}
C(t) & = C_0 \e^{\textstyle -\lambda t},  \\
P(t) &= P_0 \exp\left( \frac{\alpha}{\lambda} C_0 \left(\e^{-\lambda t}-1 \right) \right). 
\end{align}
	\end{subequations}
Then we define function $S_f$ describing the survival fraction as follows
$$S_f(t)=\frac{P(t)}{P_0}= \exp\left( \frac{\alpha}{\lambda} C_0 \left(\e^{-\lambda t}-1 \right) \right)
\exp\left( \frac{z_0}{\mu}  \left(\e^{-\mu s}-1 \right) \right). $$ 
Thus for long times $ t \gg 1/\lambda$ we obtain the formula
 \begin{equation}
 S_f = \exp \left(-\frac{\alpha}{\lambda} C_0\right). 
 \end{equation}
Survival fraction depends exponentially on the TMZ~cell kill strength $\alpha$, the effective dose $C_0$~and inversely on the time of exposure to chemotherapy $\lambda$.~This formula is similar to the linear term in the linear-quadratic model describing the effect of a~single dose of radiotherapy on cells \cite{VanderKogel}.

\label{sec:estimates}
\subsection{Motivation and rescaled model}
In the following subsection we will compute analytical estimates for the TRP as a~function of the model parameters. We will study a~broad range of chemotherapy fractionation schemes in which the time between doses is larger than the time of cell damage repair and drug uptake, \emph{i.e.} doses are separated by more than several hours.

Without loss of generality, we can assume that the time of the first drug administration is $t_1=0.$ 
In agreement with clinical practise, we will assume all drug doses to be equal. Thus $d_{j}=d$ for $j \in \{1,\ldots, n\}$ with $n$ being the total number of doses. The effective dose per fraction $C_0$ is calculated as in Sec.~\ref{sec:parameters}. 
In order to obtain analytical estimates let us introduce the following notation. 
Each chemotherapy cycle is described by three parameters: the cycle duration $T$ (measured in days), the number of doses per cycle $p$, and the interval between doses in each cycle $r$ (measured in days). Thus, the drug is given at times $t_1=0$, $t_2=r$, $t_3=2r$, $\ldots,$ $t_{p}=(p-1)r$, $t_{p+1}=T$, $t_{p+2}=T+r$, $\ldots,\ t_{n}=n/(p-1)T+(p-1)r.$
This definition allows for the description of many different chemotherapy schemes, including those in which administration of drug doses in a cycle is followed by some break (as $T$ can be greater than $pr$).
For instance, in the clinical trial described in~\cite{Kesari} patients were treated with TMZ given daily for seven weeks followed by four-week breaks. For that study we would take $T=77,\ r=1,\ p=49$ and the drug would be administered at days $t_1=0$, $t_2=1$, $t_3=2$, $\ldots,$ $t_{49}=48$, $t_{50}=77$, $t_{30}=78$ etc.

The total number of TMZ~cycles in a general fractionation scheme described above equals $n/p$ and the times of drug administration are
\begin{equation}\label{dosingtimes}
	t_j = (j-1)r + m \bigl(T - pr\bigr),
\end{equation}
where $m=\left\lfloor \frac{\textstyle j-1}{\textstyle p} \right\rfloor$ is the~number of completed chemotherapy cycles before dose $t_{j},$ $j \in \{1,\ldots, n\}.$
Then $m \in \{0,n/p -1\} $ and
\begin{equation} \label{dosingindexes}
j=pm + i,
\end{equation}
$i$ being the index of a dose within each~TMZ~cycle,  $i \in \{1, \ldots, p\}.$

From the first dose of chemotherapy drug at time $t_1=0$, the tumour growth and change in drug concentration are described by Eqs.~\eqref{ode} with initial conditions: 
\mbox{$ P(0)=P_0, \ D(0)=0, \ C(0)=C_0.$} 
For $t\in [t_j, t_{j+1})$ and after the end of chemotherapy the values of $P$, $D$ and $C$ change according to Eqs.~\eqref{ode}, while at $t=t_j$ values of functions $P,\ D$ and $C$ are as in \eqref{impulses} with $C_{j}=C_0$ for $j\in \{2,\ldots,n\}.$

Let us rescale the model variables by taking $x=P/K, \ y= D/K,$ \mbox{$z=\alpha C/\rho,$} $s=\rho t$ to get
\begin{subequations} \label{ode_scaled}
\begin{eqnarray} 
		\frac{d x}{d s}  & =  &x\left(1-x-y \right) - xz \\ 
		\frac{d y}{d s}  & = & -\frac{1}{k}y\left(1-x-y\right) + xz \\ 
		\frac{d z}{d s}  & =  & -\mu z
\end{eqnarray}
\end{subequations}
with initial conditions: $ x(0)=x_0=P_0/K, \ y(0)=0, \ z(0)=z_0=
\alpha C_0/ \rho,$
where $\mu=\lambda/\rho.$
The rescaled dose $z_0$ is imposed to be given at times \mbox{$s_1=0,$} $s_2, \ldots, s_{n}.$

\subsection{Time of response to chemotherapy}
As already mentioned, one of the main observable characteristics of the tumour response to therapy is the time to radiological progression, \emph{i.e.} the time until the tumour starts regrowing. In our mathematical framework it is the time $t_{\TRP}$ at which the total tumour mass attains its minimum, \emph{i.e.} 
$P(t_{\TRP})+D(t_{\TRP})=\displaystyle \min_{t \ge t_n}\left\{P(t)+D(t)\right\}$.
In terms of rescaled variables we look for $s_{\TRP}$ such that 
$$ x(s_{\TRP})+y(s_{\TRP})=\displaystyle \min_{s \ge s_n}\left\{x(s)+y(s)\right\}.$$ 
We will focus only on cases of tumours responding to chemotherapy with $\alpha>0$, thus showing a radiologically visible decrease in total volume. Thus, in our approach, in which only first-line chemotherapy is described and resistant cells do not arise, tumour progression will occur after the end of chemotherapy ($s_{\TRP} \gg s_{n}$) provided tumour growth is slow as happens in the case of LGGs. Thus we will try to obtain explicit formulae approximating $x$ and $y$~for $s \gg s_{n}.$

We will consider tumours with initial sizes significantly smaller than the carrying capacity, then the Eq. \eqref{ode_scaled} takes the simpler form
\begin{subequations}\label{ode_final}
\begin{eqnarray} 
		\frac{d x}{d s}  & = & x - xz, \\ 
		\frac{d y}{d s}  & = & -\frac{1}{k}y + xz, \\ 
		\frac{d z}{d s}  & =  & -\mu z,
	\end{eqnarray}
\end{subequations}
with initial conditions: $ x(0)=x_0, \ y(0)=0, \ z(0)=z_0.$ Then for rescaled time TRP we have $x'(s_{\TRP}) + y'(s_{\TRP})=0,$ therefore 
\begin{equation} \label{formula s_ttp}
x(s_{\TRP})=\frac{1}{k}y(s_{\TRP}).
\end{equation}
Eqs.~\eqref{ode_final} are a~set of ordinary differential equations with impulses, the functions $x$ and $y$ being continuous, and $z$~being discontinuous at times $s_{j}$ for $j \in \{2,\ldots,n\}$ as 
\begin{equation}
  z(s_j)=z(s_j^-)+z_0.
  \end{equation}
Since dose clearance time is about two hours we may assume that each dose is cleared in one day, then, in the rescaled units
\mbox{$z((s_{j}+\rho)^{-}) \approx 0$} for \mbox{$j \in \{1,\ldots,n\}.$} Therefore, we can approximate
\begin{equation} \label{z approximation}
z(s) \approx 
\left\{
	\begin{aligned}
		& z_0 \e^{\textstyle -\mu(s-s_{j})}  && s \in (s_{j}, s_{j}+\rho), \\ 
		& 0 && \mbox{for other $s$},
	\end{aligned}
	\right.
\end{equation} 
where $j=\displaystyle \operatorname*{arg\,max}_{i \in \{1,\ldots, n\}} \left\{ s_{i} \leq s \right\}.$
Let us define
\begin{subequations}
\begin{align}
& w(s)  =\int_{0}^{s} z(t) \dd t, \\
& w_0  =w(\rho)=\int_{0}^{\rho} z(t) \dd t= \frac{z_0}{\mu}\left(1-  \e^{\textstyle -\mu \rho}  \right).
\end{align}
\end{subequations}
We should emphasise that for $s> s_2,$ $w(s)\neq z_0\left(1-  \e^{\textstyle -\mu s}  \right)/\mu$ due to the administration of the next drug dose.
Furthermore, from the definition of function $z$ we have 
\begin{equation} \label{periodicity}
\begin{aligned}
w(s) & =  \int_{0}^{s_j} z(t) \dd t + \int_{s_{j}}^{s} z(t) \dd t =  w(s_{j}) + \int_{0}^{s-s_{j}} z(t) \dd t 
\approx (j-1)w_{0}+ w(s-s_{j}) \\
& \approx  \left\{ \begin{array}{ll}
(j-1)w_{0}+  \frac{z_0}{\mu}\left(1-  \e^{\textstyle -\mu (s-s_{j})}  \right) & s-s_{j} \leq \rho,\\
jw_0 & \textrm{otherwise,}
\end{array} \right.
\end{aligned}
\end{equation}
where $j=\displaystyle \operatorname*{arg\,max}_{i \in \{1,\ldots, n\}} \left\{ s \geq s_{i} \right\}.$
Hence for $s>s_{n}+\rho$ the formulae for rescaled proliferating and the damaged part of tumour take the~form
\begin{subequations}\label{xy}
\begin{eqnarray} \label{x}
x(s) &= & x_0 \e^{\textstyle s-w(s)}=x_0 \e ^{\textstyle s - n w_0 },  \\
\label{y}
y(s) & = & \int_{0}^{s} \e^{\textstyle -\frac{s-t}{k}} x(t) z(t) \dd t. 
\end{eqnarray}
\end{subequations}
We look for $s_{\TRP}$ which fulfils condition \eqref{formula s_ttp}. Using Eqs. \eqref{xy} we have 
\begin{equation}
k\e^{ \textstyle \tilde{k}s_{\TRP} -n w_0} = \int_{0}^{s_{\TRP}} \e^{\textstyle \tilde{k}t - w(t) } z(t) \dd t,
\end{equation}
\begin{equation} \label{sTTP1}
s_{\TRP}=\frac{1}{\tilde{k}}\Bigg[ nw_0 + \ln \left( \frac{1}{k} \int_{0}^{s_{\TRP}} \e^{\textstyle \tilde{k}t - w(t) } z(t) \dd t \right) \Bigg],
\end{equation}
where $\tilde{k}=1+1/k$. 
As a~result of approximations \eqref{z approximation} and \eqref{periodicity} we conclude that for \mbox{$s\geq s_{n}+ \rho$}  
\begin{align} \label{rhs}
& \displaystyle \int_{0}^{s} \e^{\textstyle \tilde{k}t - w(t) } z(t) \dd t =
\displaystyle \sum_{j=1}^{n} z_0 \displaystyle \int_{s_{j}}^{s_{j}+\rho} \e^{\textstyle \tilde{k}t - w(t)} \e^{\textstyle -\mu(t-s_{j})} \dd t 
\nonumber \\
 & = \displaystyle z_0 \sum_{j=1}^{n} \e^{\textstyle -(j-1)w_{0} +\tilde{k}s_{j}} \displaystyle \int_{s_{j}}^{s_{j}+\rho} \e^{\textstyle 
(\tilde{k} -\mu)(t-s_{j}) +  \frac{\textstyle z_0}{\textstyle \mu}\left(\e^{\textstyle -\mu (t-s_{j})} -1 \right)} \dd t 
\nonumber \\
& =   z_0 \left( \sum_{j=1}^{n} \e^{\textstyle -(j-1)w_{0} +\tilde{k}s_{j}} \right) \int_{0}^{\rho}
\e^{\textstyle 
(\tilde{k} -\mu)t + \frac{\textstyle z_0}{\textstyle \mu}\left( \e^{\textstyle -\mu t}-1\right)} \dd t. 
\end{align} 
Using Taylor expansion of an exponential function for $t<1/\mu$ we approximate 
\[
	\e^{-\mu t}-1 \approx \begin{cases}
			-\mu t & 0\le t<1/\mu, \\
			-1 & t\ge 1/\mu,
		\end{cases}
\]
and obtain
\begin{flalign} \label{integral}
& \displaystyle \int_{0}^{\rho}
\e^{\textstyle (\tilde{k} -\mu)t + \frac{\textstyle z_0}{\textstyle \mu}\left( \e^{\textstyle -\mu t}-1\right)} \dd t  \approx
\displaystyle \int_{0}^{\frac{1}{\mu}} \e^{\textstyle  (\tilde{k} -\mu)t - z_0 t} \dd t 
+ \int_{\frac{1}{\mu}}^{\rho} \e^{\textstyle  (\tilde{k} -\mu)t - \frac{z_0}{\mu}} \dd t \nonumber \\
&= \frac{1}{\tilde{k}-\mu -z_{0}}
\left( \e^{\textstyle \frac{\tilde{k}-\mu-z_0}{\mu}} -1 \right) +
\frac{1}{\tilde{k}-\mu} \left( \e^{\textstyle (\tilde{k}-\mu)\rho -\frac{z_0}{\mu}}-\e^{\textstyle\frac{\tilde{k}-\mu-z_0}{\mu}} \right) \nonumber \\
&= \frac{1}{\left(\tilde{k}-\mu -z_{0}\right)\left(\tilde{k}-\mu\right)} \Bigg[ z_0 \e^{\textstyle \frac{\tilde{k}-\mu-z_0}{\mu}} -\tilde{k} +\mu + 
\left(\tilde{k} -\mu -z_0\right) \e^{\textstyle \left(\tilde{k}-\mu\right)\rho - \frac{z_0}{\mu}} 
\Bigg].
\end{flalign}
To compute the sum term in Eq.~\eqref{rhs} we need the relation between dose indexes $j$~and the times of their administration $s_{j}.$ Taking into consideration assumptions from Sec. \ref{sec:estimates} and Eqs.~(\ref{dosingtimes}-\ref{dosingindexes}) we get
$  s_{j}= \big[(i-1)r+ mT\big]\rho, $
where $m\in \{0, n/p -1\} $ is the~number of completed chemotherapy cycles before dose $s_{j},$ $j \in \{1,\ldots, n\}$ and $i \in \{1, \ldots, p\}$ is an index of dose within the~TMZ~cycle.
As a~result we obtain
\begin{multline}  
\sum_{j=1}^{n} \exp \left(-(j-1)w_{0} +\tilde{k}s_{j}\right)   \\
=  \sum_{m=0}^{ n/p-1} \sum_{i=1}^{p}
\exp \left( m \left(-pw_{0} +\tilde{k}\rho T \right)+ (i-1)(-w_{0} +\tilde{k}\rho r) \right)
 \\
 = \frac{1-\e^{\textstyle \left(-p w_{0} +\tilde{k}\rho T \right) \frac{n}{p} }}{1-\e^{\textstyle -pw_{0} + \tilde{k}\rho T}} 
\cdot \frac{1-\e^{\textstyle \left(-w_{0} +\tilde{k}\rho r \right)p}}{1-\e^{\textstyle -w_{0} +\tilde{k}\rho r}},
 \end{multline}
where $n$ is the total number of doses and $n/p$ is the total number of chemotherapy cycles as previously stated. 
Then using Eqs. \eqref{sTTP1}, \eqref{rhs} and \eqref{integral} we get 
\begin{align}
& s_{\TRP}  = 
\frac{n w_0 }{\tilde{k}}
+ \frac{1}{\tilde{k}}\ln 
\left\{ z_0 \e^{\textstyle \frac{\tilde{k}-\mu-z_0}{\mu}} -\tilde{k} +\mu + 
\left(\tilde{k} -\mu -z_0\right) \e^{\textstyle \left(\tilde{k}-\mu\right)\rho  - \frac{ z_0}{\mu}} 
\right\} \nonumber \\
& + \frac{1}{\tilde{k}}\ln \left\{
\frac{\textstyle z_0\left( 1-\e^{\textstyle \left(-pw_{0} +\tilde{k}\rho T \right) \frac{n}{p} }\right) 
\left(1-\e^{\textstyle \left(-w_{0} +\tilde{k}\rho r\right)p}\right)}
{\textstyle k\left( 1-\e^{\textstyle -pw_{0} + \tilde{k}\rho T} \right) 
\left(1-\e^{\textstyle -w_{0} +\tilde{k}\rho r}\right) 
\left(\tilde{k}-\mu -z_{0} \right)\left(\tilde{k}-\mu\right)} \right\} .
\end{align}
In terms of the initial time scale, the time to tumour progression can be estimated as 
$t_{\TRP}=s_{\TRP}/ \rho,$ giving the final result
\begin{align} \label{formula}
& t_{\TRP}  = 
\frac{n w_0 }{\tilde{k}\rho}
+ \frac{1}{\tilde{k}\rho}\ln 
\left\{ z_0 \e^{\textstyle \frac{\tilde{k}-\mu-z_0}{\mu}} -\tilde{k} +\mu + 
\left(\tilde{k} -\mu -z_0\right) \e^{\textstyle \left(\tilde{k}-\mu\right)\rho  - \frac{ z_0}{\mu}} 
\right\} \nonumber \\
& + \frac{1}{\tilde{k}\rho}\ln \left\{
\frac{\textstyle z_0\left( 1-\e^{\textstyle \left(-pw_{0} +\tilde{k}\rho T \right) \frac{n}{p} }\right) 
\left(1-\e^{\textstyle \left(-w_{0} +\tilde{k}\rho r\right)p}\right)}
{\textstyle k\left( 1-\e^{\textstyle -pw_{0} + \tilde{k}\rho T} \right) 
\left(1-\e^{\textstyle -w_{0} +\tilde{k}\rho r}\right) 
\left(\tilde{k}-\mu -z_{0} \right)\left(\tilde{k}-\mu\right)} \right\}.
\end{align}
Eq. \eqref{formula} gives the time to radiological progression as a~function of parameters with relevant biological and/or therapeutical meaning.
Since TRP is a~metric of practical relevance, it is very interesting that it is possible to estimate its value analytically. 

\subsection{Validation}
Eq. \eqref{formula} has been obtained via a~number of approximations and thus it is relevant to compare its predictions with the results of the original equation \eqref{ode} and real patient data.

To do the latter we first fix the treatment parameters to match those routinely used for TMZ therapeutic schedules. Since TMZ~is given on 5 consecutive days in cycles consisting of 28 days, we get $p=5$, $T=28$, $r=1$. Taking dose per fraction to be 150~mg/m$^2$ and the other parameters as in Sec.~\ref{sec:estimates} we can then estimate the time of progression for the individual patients studied previously (accounting for the number of cycles received by each patient). 

\begin{figure}[h!p]
\centering
\includegraphics[width=0.525\textwidth]{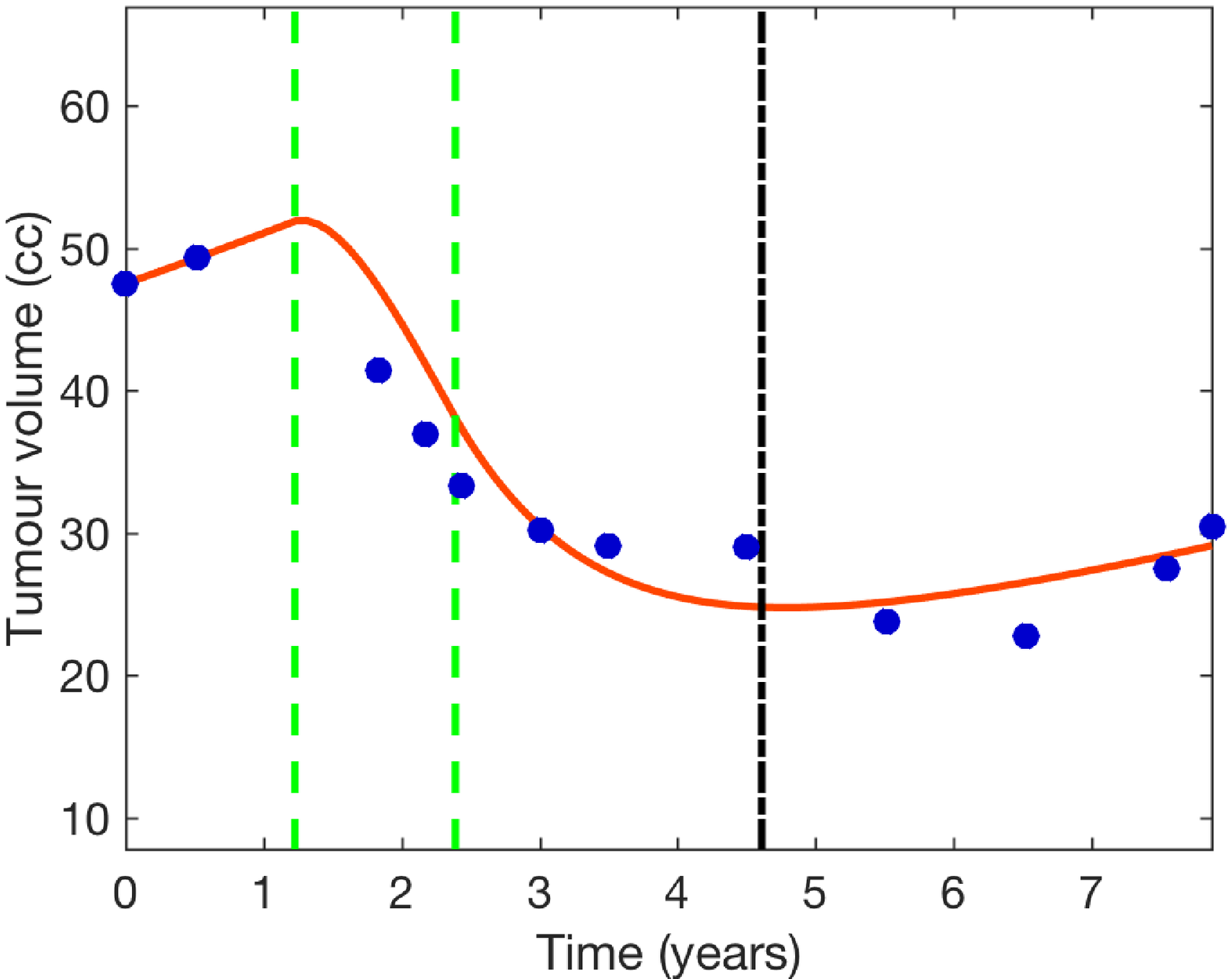}
\includegraphics[width=0.525\textwidth]{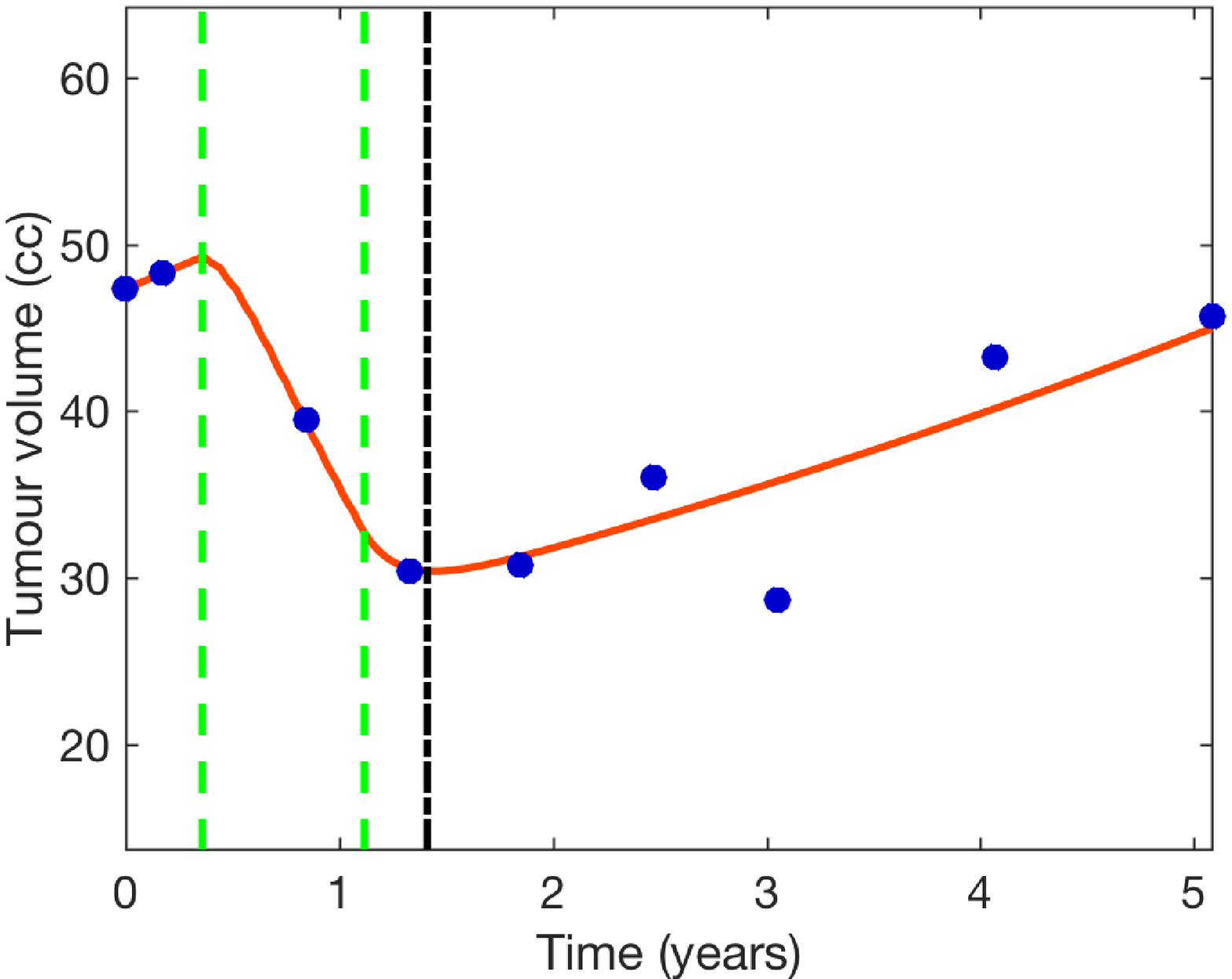}
\includegraphics[width=0.525\textwidth]{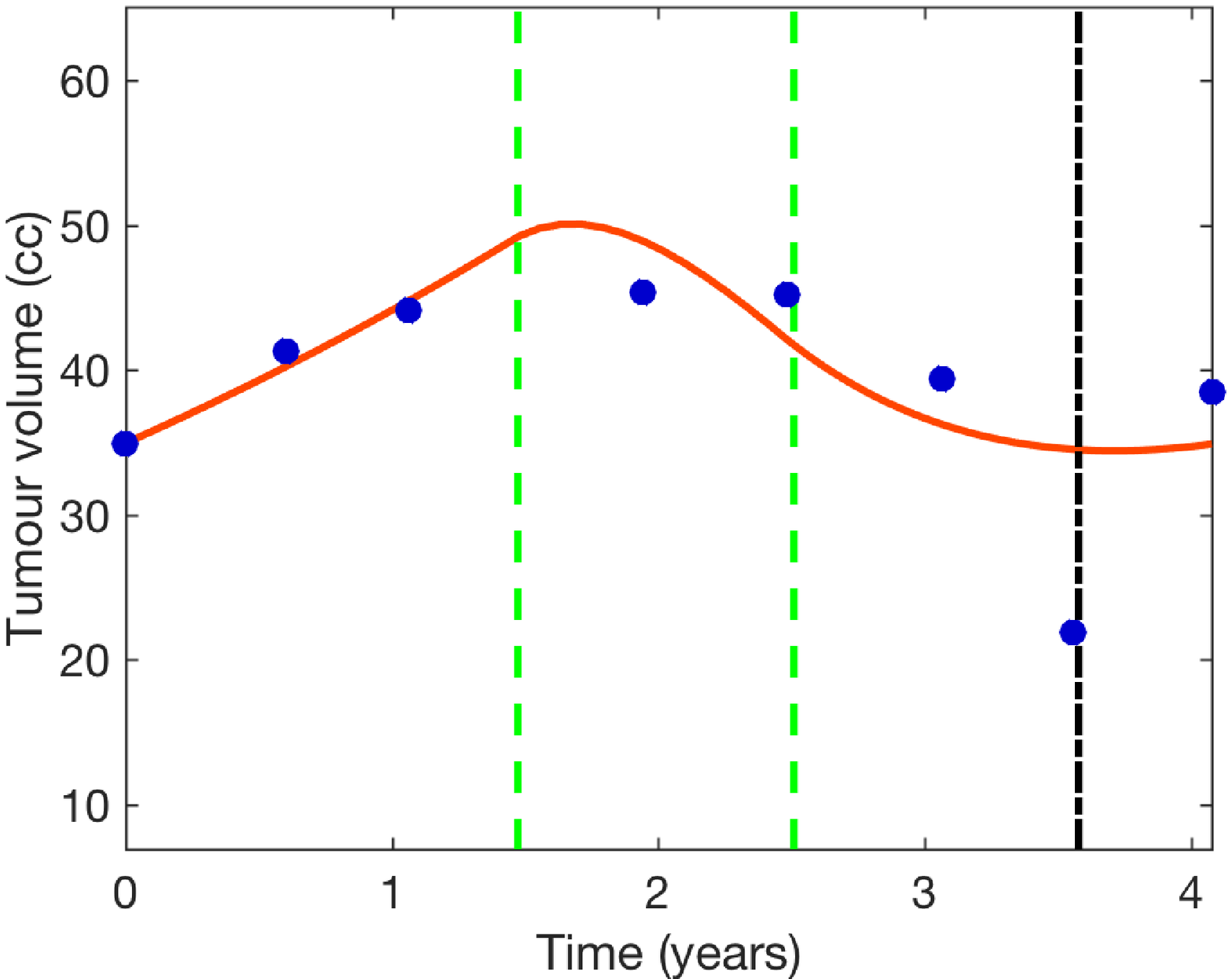}
\caption{Tumour volume evolution for three  patients treated with TMZ. 
Vertical dashed lines mark the start and the end of TMZ~treatment. 
Circles denote the volumes obtained from MRIs and solid lines the results of the best fit using Eqs.~\eqref{ode}. 
(top) Woman treated with 6~TMZ cycles,  $\alpha=0.199094$ml/$\mu$g/day, $\rho = 0.00022/$day, $k = 0.075644.$
(center) Man treated with  11~TMZ~cycles, $\alpha = 0.17367$ml/$\mu$g/day, $\rho = 0.000338/$day, $k = 0.019279.$ 
(bottom) Man treated with 4~TMZ~cycles, $\alpha=0.236439$ml/$\mu$g/day, $\rho =  0.000701/$day, $k = 0.257806.$
The times to radiological progression computed using Eq.~\eqref{formula} are marked with vertical dashed-dotted lines, showing a very good agreement with the data and the simulations of Eqs.~\eqref{ode}. The relative differences between $t_{\TRP}$ calculated from Eq.~\eqref{formula} and from Eqs.~\eqref{ode} were respectively $0.042043, \ 0.041671$ and $0.037481$ years, for the selected patients. 
}
\label{fig:tTTP}
\end{figure}

Fig.~\ref{fig:tTTP}~shows how well formula \eqref{formula} estimates the response to chemotherapy for three patients chosen from our database. The task of comparing simulation results with real patients MRI data requires a lot of caution. In particular, we need to take into account the limitations of calculations of tumour volume using the method of three largest diameters. The method is only an approximation of real tumour volume and its accuracy is limited by slice thickness, changes in head position \cite{Schmitt} or even by perception of medical doctor who calculate these diameters. However in this case one can conclude that we have obtained a satisfactory result in fit. In the future we hope that MRI data will be analysed through automatic segmentation, \eg~with algorithm suggested by Porz \etal \cite{Porz} and the real tumour volume will be calculated more accurately. 

\begin{figure}[h!p]
	\begin{center} 
		\includegraphics[trim={0.6cm 0.1cm 1.4cm 0.9cm},clip=true,width=0.49\textwidth]{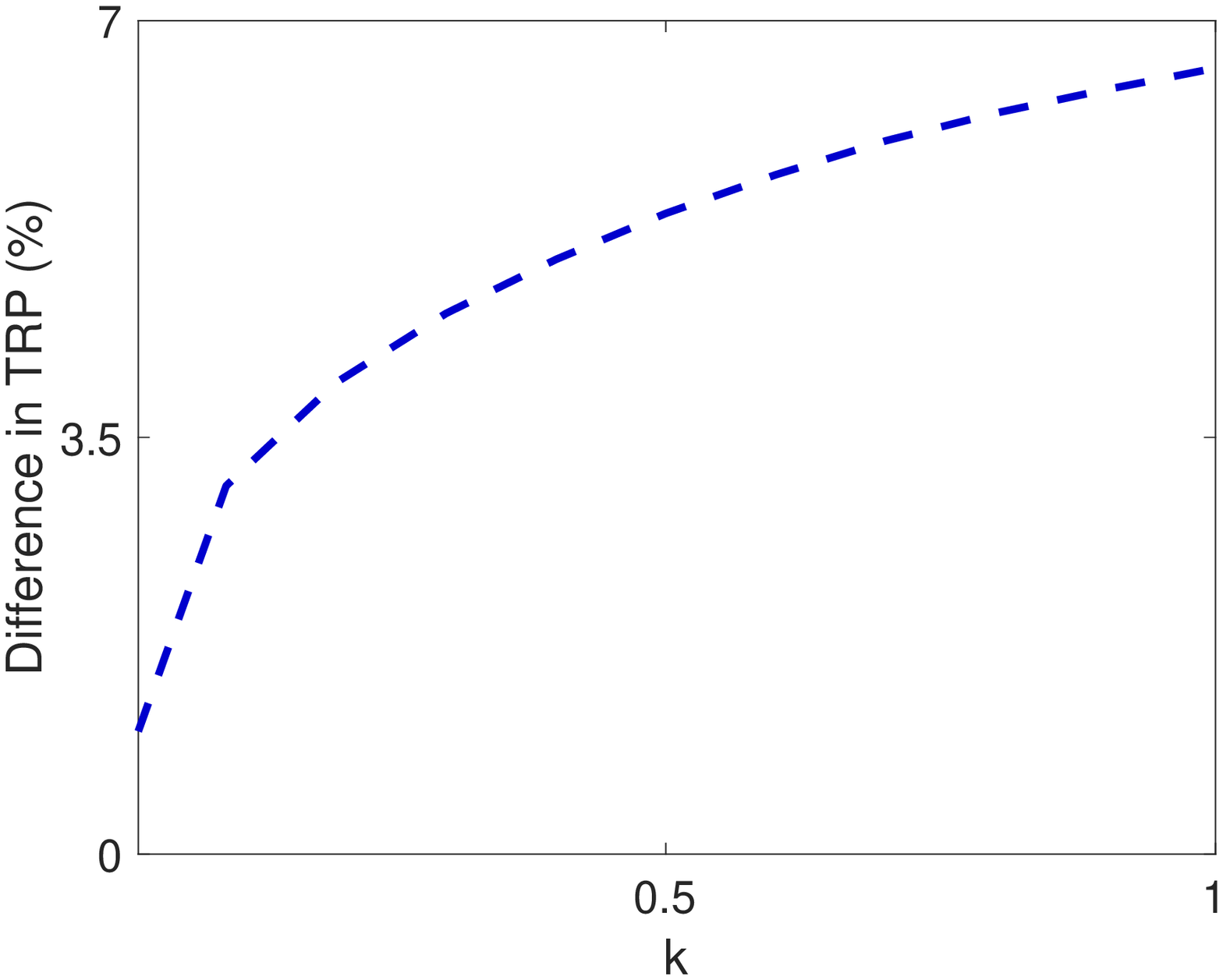}
		\includegraphics[trim={0.6cm 0.1cm 1.4cm 0.9cm},clip=true,width=0.49\textwidth]{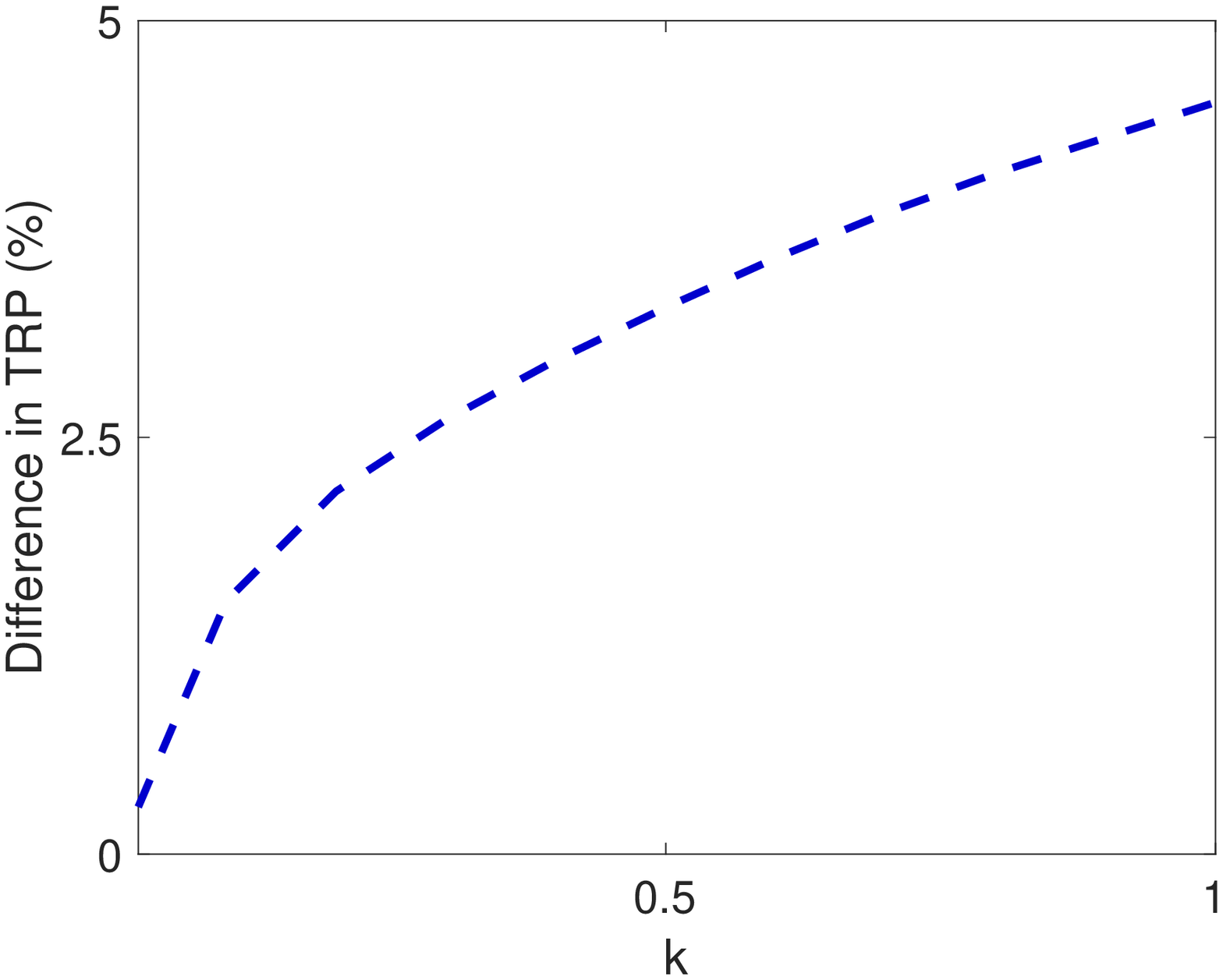}
		\ \\ \ \\		
		\includegraphics[trim={0.6cm 0.1cm 1.4cm 0.9cm},clip=true,width=0.49\textwidth, height=0.2\textheight]{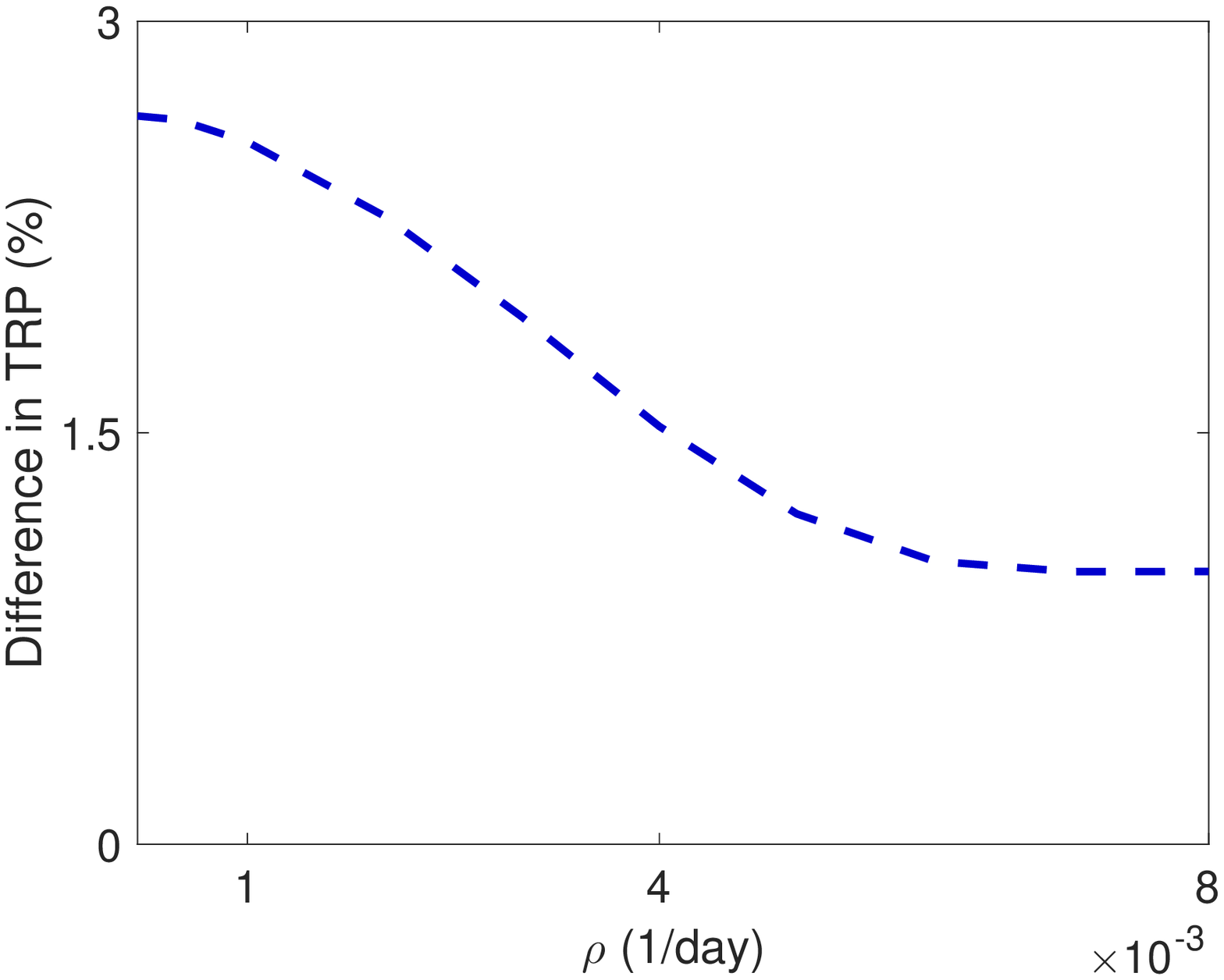}		
		\includegraphics[trim={0.6cm 0.1cm 1.4cm 0.9cm},clip=true,width=0.49\textwidth, height=0.2\textheight]{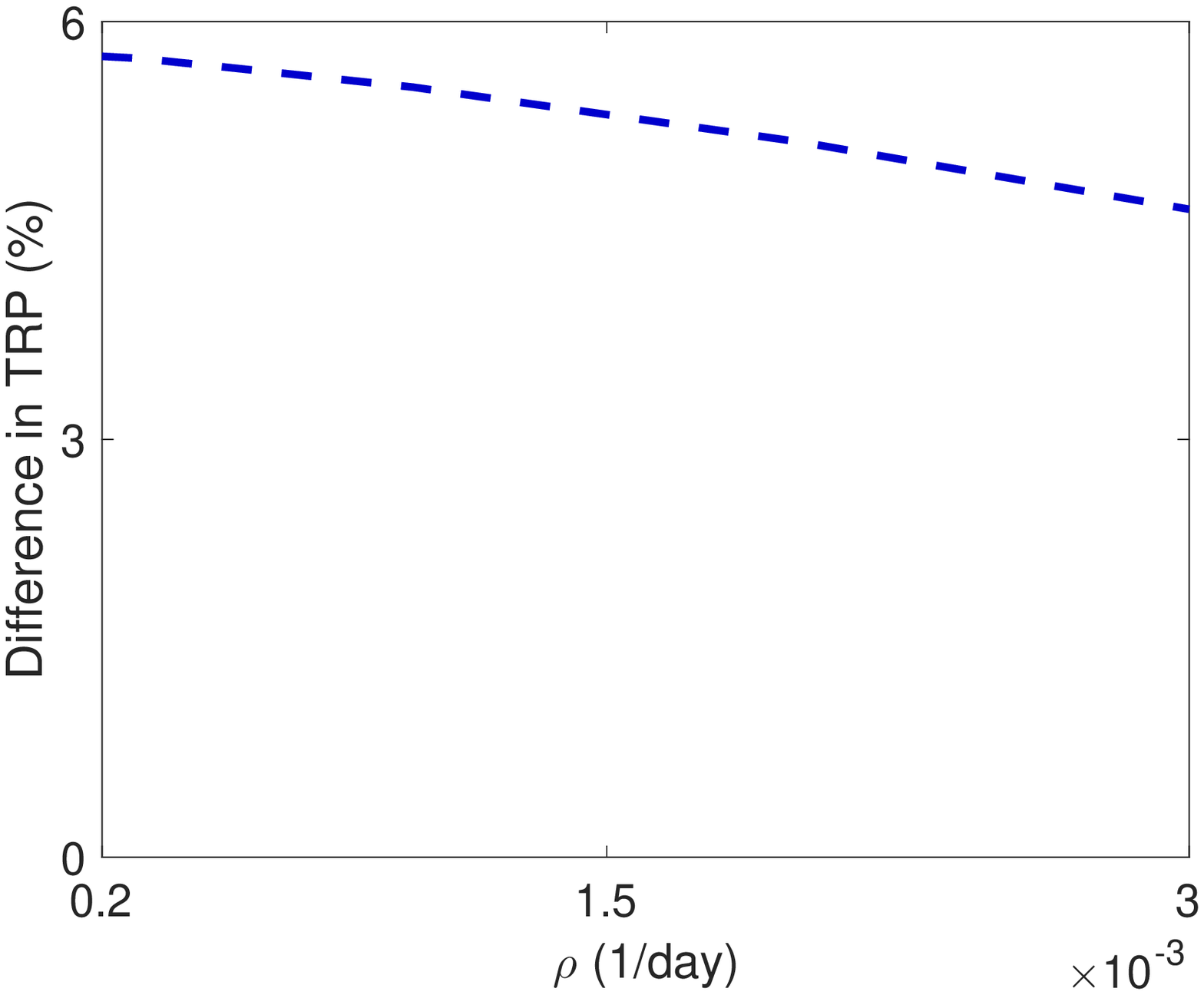}
		\ \\ \ \\	
		\includegraphics[trim={0.6cm 0cm 1.4cm 0.9cm},clip=true,width=0.49\textwidth, height=0.2\textheight]{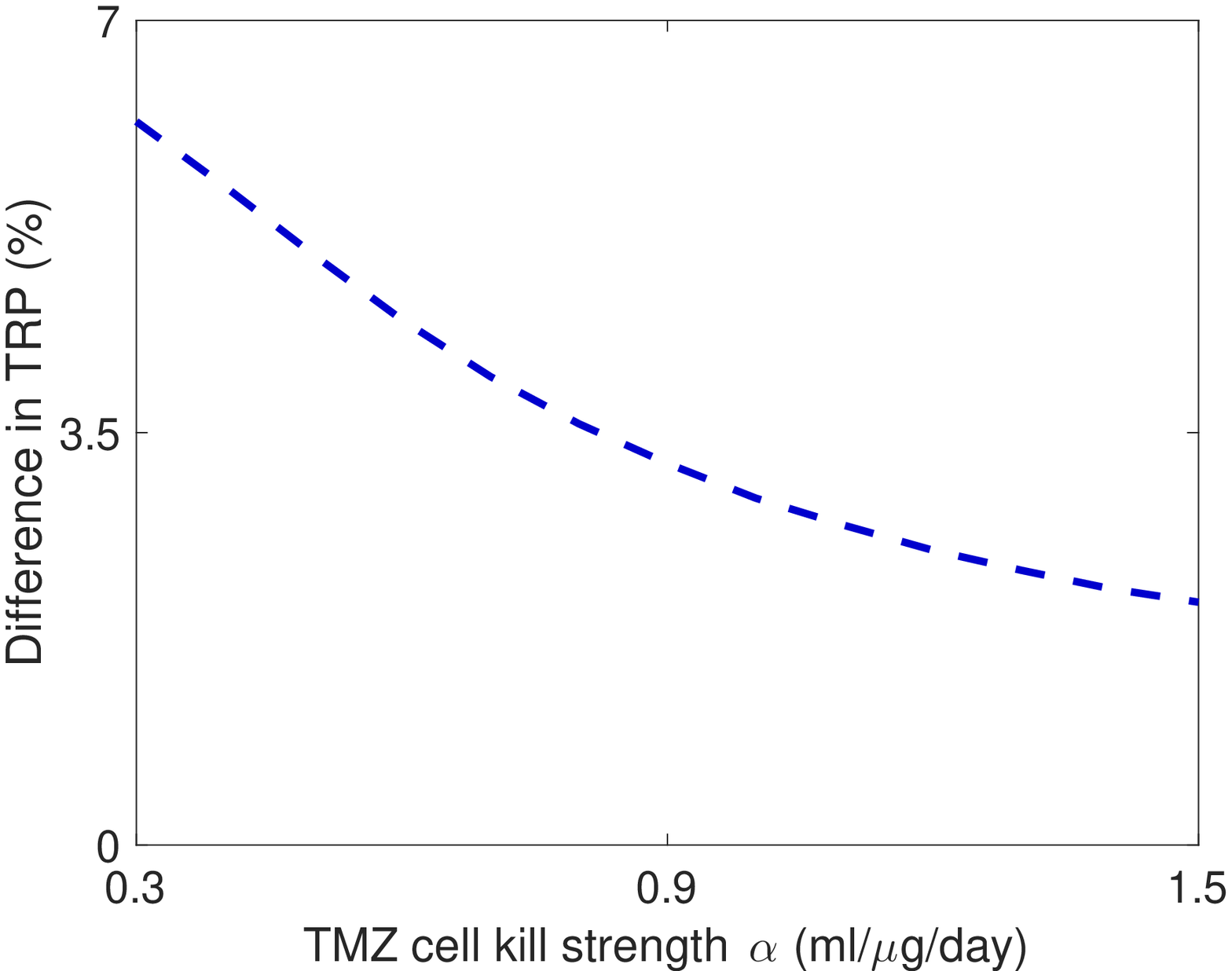}
		\includegraphics[trim={0.6cm 0cm 1.4cm 0.9cm},clip=true,width=0.49\textwidth, height=0.2\textheight]{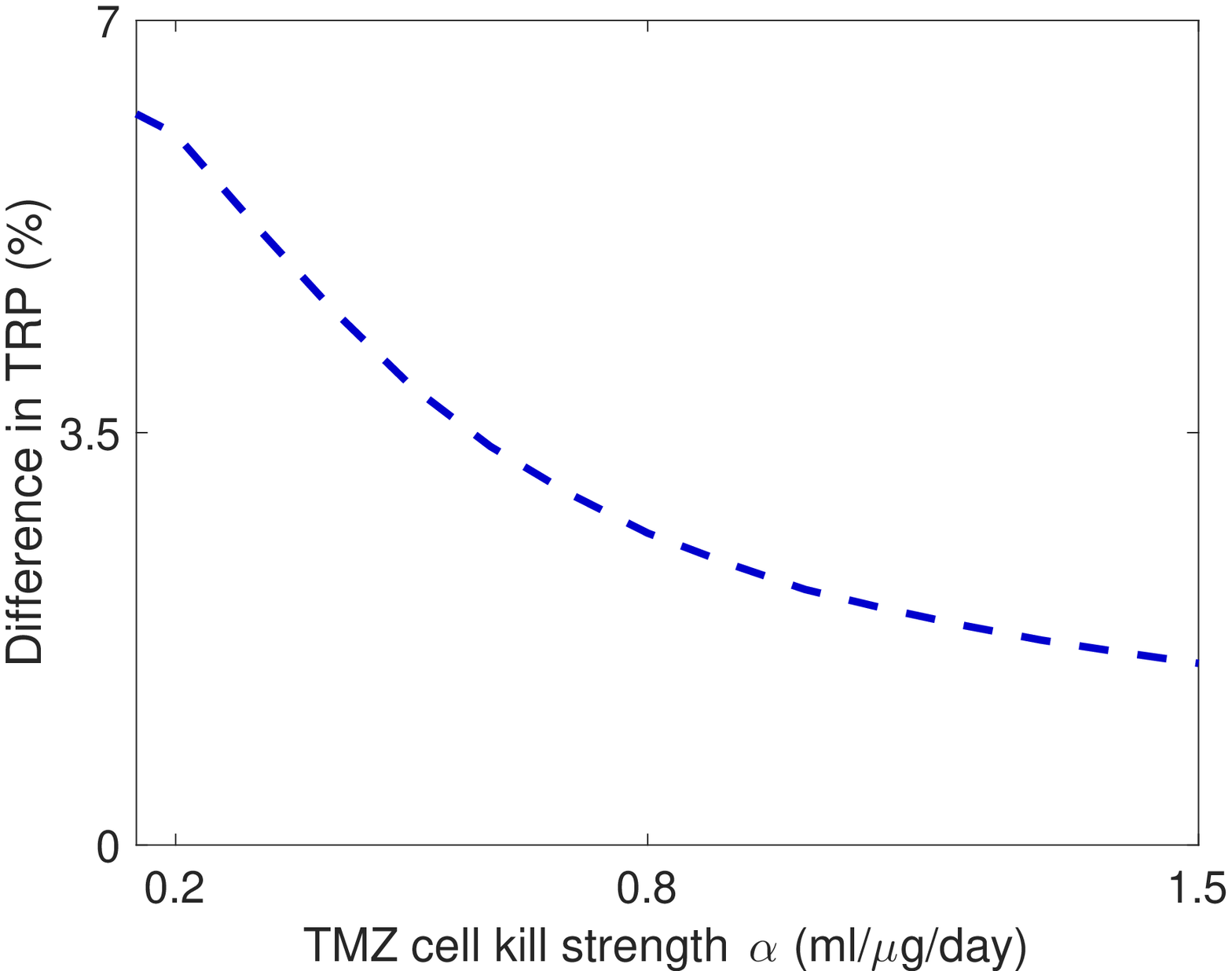}
		\caption{Relative percentage difference between time to radiological progression estimated from simulations of model~\eqref{ode} and formula \eqref{formula}. We considered 12 cycles of TMZ~as in the standard fractionation scheme (see Sec.~\ref{sec:parameters}) for virtual patients with LGG having an initial volume of 40 cm$^3$. Results for $\rho = 0.0004$/day, $\alpha=0.4$ml/$\mu$g/day, $k \in [0.02, 1]$ (left) and $\rho = 0.0008$/day, $\alpha=0.8$ml/$\mu$g/day, $k \in [0.02, 1].$ (center) Results for $\alpha=0.8$ml/$\mu$g/day, $k=0.3$ for $\rho \in [0.2,8] \times 10^{-3}$/day (left) and $\alpha=0.8$ml/$\mu$g/day, $k=0.3$ for $\rho \in [0.2,3] \times 10^{-3}$/day. (bottom) Results for $\rho = 0.0008$/day, $k=0.6$ and $\alpha \in [0.3,1.5]$ml/$\mu$g/day (left) and $\rho = 0.0004$/day, $k=0.3$ and $\alpha \in [0.15,1.5]$ml/$\mu$g/day (right).
		}
		\label{fig:diff}
	\end{center}
\end{figure}

Fig.~\ref{fig:diff} presents relative differences between TRP from the estimated formula and simulations for different sets of parameters, suggesting a very good approximation.

\subsection{The study of tumour response for other chemotherapy protocols}
\begin{figure}[h!t]
\begin{center}
\includegraphics[trim={0.8cm 0cm 1.8cm 0.9cm},clip=true,width=0.46\textwidth]{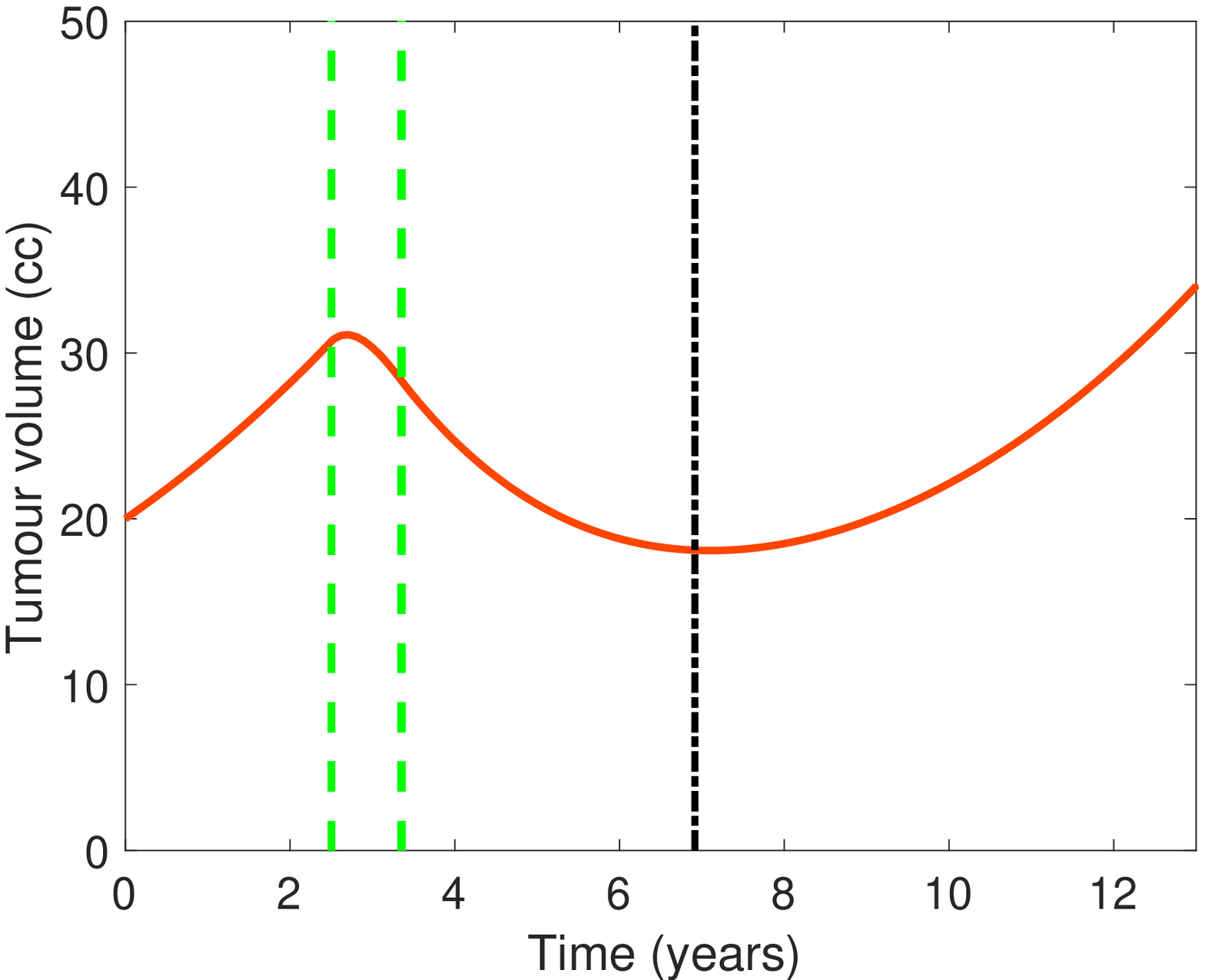}
\includegraphics[trim={0.8cm 0cm 1.8cm 0.9cm},clip=true,width=0.46\textwidth]{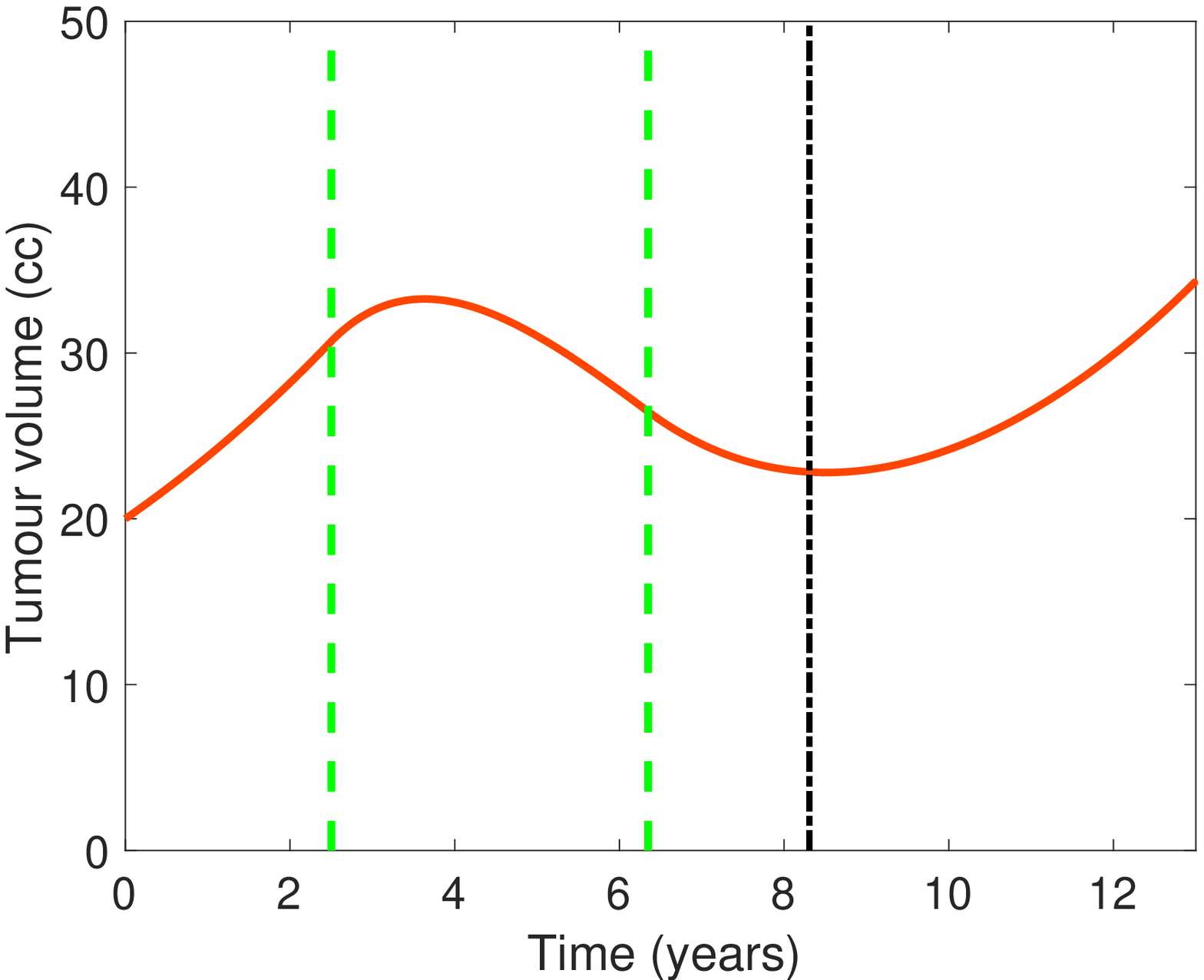}
\caption{Tumour volume evolution for virtual patients simulated from Eqs.~\eqref{ode}. Times to radiological progression estimated from Eq.~\eqref{formula} are marked with vertical dashed-dotted lines. Values of parameters were $k =0.5$, $\alpha = 0.4$ml/$\mu$g/day and $\rho = 0.005/$day. The start and the end of TMZ~treatment are marked with vertical dashed lines. (left) 9 TMZ~cycles of 34 days with  doses given every 2 days for a~total of 10 doses per cycle. The dose per fraction was $d$=100~mg/m$^2.$ (right) 18~TMZ~cycles of 77 days with doses given every 7 days for a~total of 10 doses per cycle. The dose per fraction was $d$=50 mg/m$^2.$ Relative differences between times free of progression calculated from Eq.~\eqref{formula} and estimated from simulations were 0.026756 and 0.025303 years, respectively.}
\label{fig:tTTP1}
\end{center}
\end{figure}

We have also verified that Eq.~\eqref{formula} provides a good approximation of TRP for model \eqref{ode} for other fractionation schemes. Fig.~\ref{fig:tTTP1} shows some examples. 

\section{Discussion and therapeutic implications}

Due to the observed clinical significance of TMZ there is an increasing interest in studying its characteristics. Up to now there have been a great deal of relevant research into the pharmacokinetic/pharmacodynamic properties of TMZ \cite{Baker,Hammond,Ostermann,Portnow}, its specific mechanism of action \cite{Marchesi,Roos,Barciszewska,Agarwala} and modelling its concentration dynamics \emph{in vitro} and \emph{in vivo} \cite{Zhou,Rosso,Ballesta}. However there are fewer mathematical studies of patient response to TMZ in LGGs. 
Here we intended to construct a mathematical model which would enable understanding of delayed response to chemotherapy observed in LGGs without using an excessive number of unknown parameters. Cells were assumed to grow logistically, chemotherapy drug kinetics and its effect on glioma cells was based on TMZ concentration in brain tissue \cite{Portnow} and clinical observations \cite{Bent1,Ricard,Chamberlain}. Note that even the authors of the very complicated model \cite{Ballesta}, constructed for the purpose of describing pharmacokinetics and pharmacodynamis of TMZ, validated their model for human cerebral tumours on the basis of data of Portnow \etal \cite{Portnow}.
It is remarkable that a~simple model such as the one presented here with essentially only three unknown parameters ($\alpha, \rho, k$) is able to describe the response of real patients to a~variable number of cycles of TMZ.

The model also shows a~correlation between a~short time to radiological progression and a~poor virtual patient outcome. We may conclude that time to radiological progression can be useful as a~measure of tumour aggressiveness due to its dependence on tumour-specific parameters: proliferation rate $\rho$ and TMZ~cell kill strength $\alpha$ (see Figs.~\ref{fig:relacje},\ref{fig:surf}). 
Our data on patients treated with first-line TMZ~suggests likewise that despite other therapies used in the follow-up, patients who 
had shorter estimated TRP had worse prognosis. Such observation has been made for radiotherapy \cite{Pallud1, Ducray}, but so far no similar analysis of response to TMZ~has been done. The velocity of tumour decrease after radiotherapy (or equivalently time of progression-free survival) is strongly associated with the risk of rapid progression and poor overall survival. Here we suggest a similar result for the response to chemotherapy, namely that short time to radiological progression results in shorter overall survival. 

This outcome makes us think of the possibility of using chemo\-therapy to probe tumours, hence providing estimates of tumour-specific parameters $\rho$ and $\alpha.$ We could apply a small number of cycles of TMZ~causing minimal toxicity and monitor with MRI~the tumour response to chemotherapy. In order to assure the reasonable measurement error, we would need at least two measurements before and three  after TMZ onset. We predict that the time horizon would be of around 2 years from the time of the first MRI. We believe it could be feasible as even up to now there were cases when MRI was performed three times a year. Based on our database there will be no progression at this time horizon. Such a procedure can be used as a~novel way to assess tumour aggressiveness. Our mathematical model suggests that tumour which attains its minimal volume early after a short course of TMZ~treatment (has shorter TRP)~may be more aggressive, therefore in such a~case the remaining TMZ~dose has to be finished as soon as possible and other therapeutic options (further surgery if feasible or radiotherapy) should be considered. Such a concept can be supported also by the\emph{in vitro} results of Roos \etal \cite{Roos}, who shown that higher proliferation rates accelerate apoptosis after TMZ treatment, thus in terms of our idea, shorter TRP.

This idea resembles that described in \cite{Victor}, but with chemotherapy instead of radiotherapy.~Although the modelling principles are similar, from the clinical point of view the use of TMZ~is a much more interesting as a way to probe a tumour than the use of radiotherapy as its side-effects are long-term and non-reversible. On the other hand, TMZ~has significantly lower and largely reversible side-effects. Moreover, radiotherapy is well known to induce changes in the MRI~images due to inflammation which may distort the analysis of the tumour response. Finally, TMZ~is easily managed because it is administered orally.

\section{Conclusions}
We have build a~model which is simple from the mathematical point of view, but which incorporates the basic biological features of LGG growth and the response to chemotherapy. 

The model is able to describe response to TMZ with a minimal number of parameters and suggests that tumours having a shorter time to radiological response after TMZ~treatment~may be more aggressive in terms of proliferation potential.~We plan to reassess this observation using a~larger data set, if possible.~In this case we would like also to verify whether the survival curves obtained can be fitted for a~cohort of virtual patients, as done by Kirkby \etal \cite{Kirkby2007}.  

Moreover, we propose a~paradigm for probing tumour with TMZ~which could be used in clinical practice. We have also found an equation giving the time free of progression as a~function of the relevant biological and therapeutic parameters. In future studies it may be helpful in designing treatment schedules giving the longest TRP possible with the additional condition for the toxicity level.

In order to address other clinically relevant issues and the biological perspective several improvements to the present model will be implemented in the near future. First it may be appropriate to include more biological details such as the potential existence of the so-called cancer initiating cells or cancer stem cells. Also incorporating different phenotypes may be relevant to describe the process of acquiring resistances to TMZ.

It would also be interesting to find the minimal doses and minimal frequency of therapy such that the solution (namely tumour mass) stays below a~given threshold. In principle, survival could be improved and chemoresistance deferred using metronomic fractionations, \emph{i.e.} schedules consisting of many, equally spaced and generally low doses of chemotherapeutic drugs without extended rest periods (see \eg~\cite{Benzekry,Andre}).

We hope that optimized cancer treatment protocols on the basis of models such as the one presented in this paper may become in the future a~standard element of personalised medicine.
	
\bibliography{article_chemotherapyLGGs4}

\end{document}